\documentclass[letterpaper,aps,prd,twocolumn,superscriptaddress,showpacs,preprintnumbers,footinbib]{revtex4}

\usepackage{graphicx}
\usepackage[intlimits,centertags]{amsmath}
\usepackage{amssymb,amsfonts}
\usepackage{gensymb}
\usepackage{mathrsfs}

\usepackage{hyperref}
\hypersetup{pdfauthor=Markus Ahlers and Kohta Murase,pdftitle=Multi-Messenger Tests for the Galactic Origin of the IceCube Excess,colorlinks=true,linkcolor=blue,citecolor=blue,filecolor=blue,urlcolor=blue}

\begin{document}

\title{Probing the Galactic Origin of the IceCube Excess with Gamma-Rays}

\author{Markus~Ahlers} 
\affiliation{Wisconsin IceCube Particle Astrophysics Center (WIPAC) and Department of Physics,\\ University of Wisconsin--Madison, Madison, WI 53706, USA}

\author{Kohta Murase}
\affiliation{Hubble Fellow -- Institute for Advanced Study, 1 Einstein Dr.~Princeton, NJ 08540, USA}

\begin{abstract}
The IceCube Collaboration has recently reported evidence for a high-energy extraterrestrial neutrino flux. During two years of operation 28 events with energies between $30$~TeV and $1.2$~PeV were observed while only $10.6$ events were expected from conventional atmospheric backgrounds. The hadronic interactions responsible for this IceCube excess will also produce a flux of high-energy $\gamma$-rays that can serve as a probe of source direction and distance. We show that existing TeV to PeV diffuse $\gamma$-ray limits support the interpretation that the IceCube excess is mostly of extragalactic origin. However, we point out that $\gamma$-ray surveys are biased in the Northern Hemisphere whereas the recent IceCube data tentatively show a weak preference for the Southern Sky. Possible sub-dominant contributions from Galactic neutrino sources like remnants of supernovae and hypernovae are marginally consistent with present $\gamma$-ray limits. This emphasizes the importance of future diffuse TeV to PeV $\gamma$-ray surveys in the Southern Hemisphere, particularly in the extended region around the Galactic Center including the {\it Fermi Bubbles}.
\end{abstract}


\preprint{}

\maketitle

\section{Introduction}

Astrophysical neutrinos serve as one of the few agents to uncover the origin and distribution of cosmic rays (CRs) in our Universe. During the acceleration in their sources or during their propagation through the interstellar or intergalactic medium CRs can undergo hadronic interactions with matter and radiation. The secondary mesons (mostly pions) of these interactions decay and produce a flux of high-energy neutrinos. Due to their feeble interactions with matter and radiation these neutrinos can probe CR interactions over a wide energy and distance range. However, this property is also the main challenge for their detection on Earth. One experimental realization consists of the detection of optical Cherenkov light emission from charged particles produced in neutrino-nucleon interactions in large volumes of transparent media that are sufficiently shielded from atmospheric background. This method has been successfully applied in Lake Baikal, the Mediterranean (ANTARES) and the Antarctic glacier (AMANDA \& IceCube) (for a review see Ref.~\cite{Halzen:2010yj}).

The IceCube neutrino observatory located at the geographic South Pole is presently the most sensitive instrument to uncover astrophysical neutrino sources in the TeV to PeV energy range. In a recent paper~\cite{Aartsen:2013pza} the IceCube Collaboration reported evidence for a flux of extraterrestrial neutrinos at the $4\sigma$ confidence level. This analysis follows up on the observation of two PeV neutrino cascades within two years of operation with the 79 and 86 string configurations~\cite{Aartsen:2013bka}. The new data set consists of 26 additional events extending the {\it observed} energy range down to 30~TeV and contains 7 tracks and 19 cascades. In comparison to the expected number of $10.6^{+5.0}_{-3.6}$ events from atmospheric muons and neutrinos, the combined observation of 2+26 events corresponds to an excess with an significance of $4.1\sigma$, increasing to $4.8\sigma$ in an {\it a posteriori} test of all 28 events~\cite{Aartsen:2013pza}. This excess of events (denoted ``IceCube excess'' in the following) is consistent with a diffuse and equal-flavor $E^{-2}$-flux at the level of
\begin{equation}\label{eq:ICnu}
E_\nu^2 J^{\rm IC}_{\nu_\alpha} \simeq (1.2\pm0.4)\times10^{-8}{\rm GeV}{\rm cm}^{-2}{\rm s}^{-1}{\rm sr}^{-1}\,,
\end{equation}
based on 17 events in the 60~TeV to 2~PeV energy range.

%%%
\begin{figure*}[t]\centering
\includegraphics[width=0.8\linewidth]{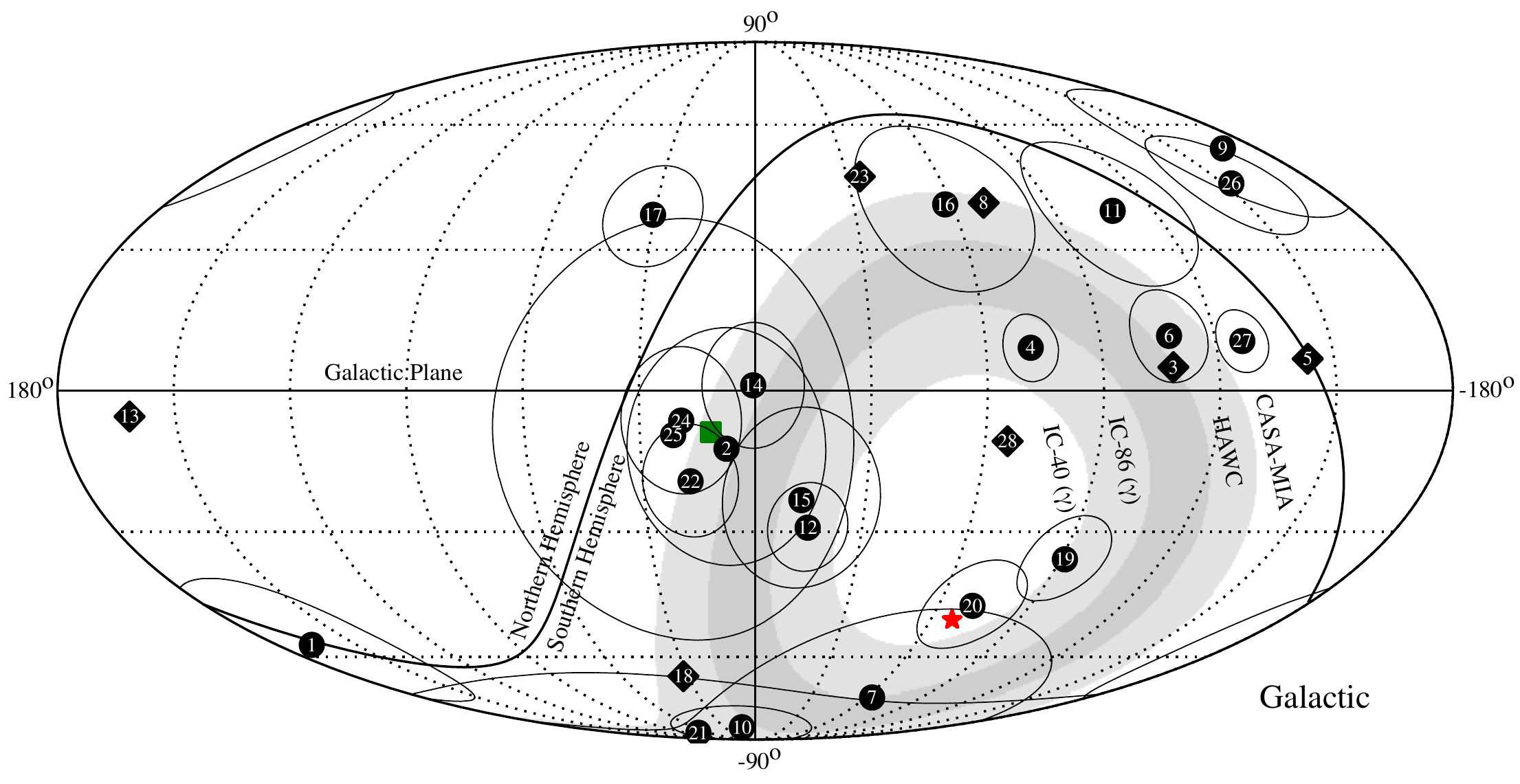}
\caption[]{Mollweide projection of the 21 cascades (circles) and 7 muons (diamonds) with time-ordered event numbers according to Ref.~\cite{Aartsen:2013pza}. The green square and red star indicate the hot-spot in the cascade clustering search and the hottest spot (with low significance) in the PeV $\gamma$-ray search~\cite{Aartsen:2012gka}. The light-gray shaded area is the region in the sky which is presently uncharted in PeV $\gamma$-ray emission. The northern and southern edge of this unaccessible region is given by the reach of CASA-MIA~\cite{Chantell:1997gs} and IceCube's $\gamma$-ray search with the IC-40 configuration~\cite{Aartsen:2012gka}, respectively. The dark-gray shaded area shows this region assuming future observations with the full IC-86 configuration and HAWC~\cite{DeYoung:2012mj}.}\label{fig1}
\end{figure*}
%%%

The distribution of the 28 neutrino events should provide additional clues about candidate sources. Utilizing the wave-form of the Cherenkov light signal in individual digital optical modules the IceCube Collaboration managed to reconstruct the initial neutrino direction of the cascade events with a median resolution of $10^\circ$ to $15^\circ$ depending on energy. In Fig.~\ref{fig1} we show the 21 cascades (filled circles) and 7 muons (diamonds) at their best-fit location and indicate the angular uncertainty (statistical and systematic) via circles. The event numbers are time-ordered using the information of Table I in Ref.~\cite{Aartsen:2013pza}. IceCube's statistical test for event clusters did not show a significant excess over an isotropic distribution of events. For the 21 cascade events the direction with the highest likelihood to be from a point source with a trial-corrected significance of 8\% (denoted ``hot-spot'' in the following and indicated as a square in Fig.~\ref{fig1}) is slightly off the Galactic Center (GC). Five cascades (2, 14, 22, 24 \& 25) overlap with the hot-spot, including one of the PeV cascades (14). These events could be associated with multiple unresolved sources or an extended source close to the GC. However, the other PeV event (20) and several other sub-PeV events do not originate from the extended region around the GC.

The origin of the IceCube excess is unknown. Standard fluxes of the prompt neutrino background~\cite{Enberg:2008te,Gaisser:2012zz,Gaisser:2013ira} from the decay of charmed mesons are too low to have a significant contribution at PeV energies, though model uncertainties have to be carefully accounted for~\cite{Lipari:2013taa}. Early studies based on the preliminary energy uncertainty of the two PeV events suggested a possible connection to the Glashow resonance~\cite{Bhattacharya:2011qu}, but the result of a dedicated follow-up analysis disfavors this possibility~\cite{Aartsen:2013bka}. In fact, the non-observation of events beyond 2~PeV suggests a break or an exponential cutoff in the flux~\cite{Laha:2013lka,Aartsen:2013pza}. Implications of the preliminary IceCube results on Galactic and extragalactic cosmic rays have been discussed in Refs.~\cite{Cholis:2012kq,Kistler:2013my,Anchordoqui:2013lna,He:2013zpa,Winter:2013cla,Chen:2013dza}.  A connection to cosmogenic neutrinos produced via the extragalactic background light seems unlikely~\cite{Laha:2013lka,Roulet:2012rv}, unless one assumes the optimistic extragalactic background light disfavored by Fermi observations of $\gamma$-ray bursts along with relatively low maximum proton energies~\cite{Kalashev:2013vba}. Various PeV neutrino sources including $\gamma$-ray bursts, peculiar supernovae, newly-born pulsars, active galactic nuclei, star-forming galaxies and intergalactic shocks have already been suggested before the discovery of the IceCube excess. In particular, the observation can be associated with extragalactic sources, {\it e.g.}, low-power $\gamma$-ray bursts~\cite{Murase:2013ffa}, cores of active galactic nuclei~\cite{Stecker:2013fxa}, star-forming galaxies~\cite{Murase:2013rfa,He:2013cqa}, intergalactic shocks and active galaxies embedded in structured regions~\cite{Murase:2013rfa}. In addition, Galactic neutrino sources have been discussed, pointing out a possible association with unidentified TeV $\gamma$-ray sources~\cite{Fox:2013oza} or the sub-TeV diffuse Galactic $\gamma$-ray emission~\cite{Neronov:2013lza}. More exotic models like the PeV dark matter (DM) decay scenario have also been suggested~\cite{Feldstein:2013kka,Esmaili:2013gha,Bai:2013nga}.

Neutrino production at TeV to PeV energies is thought to proceed via pion production via proton-photon ($p\gamma$) or proton-gas ($pp$) interactions with an inelasticity $\kappa_p$ of about 20\% and 50\%, respectively. Each of the three neutrinos from the decay chain $\pi^+\to\mu^+\nu_\mu$ and $\mu^+\to e^+\nu_e\bar\nu_\mu$ carries about one quarter of the pion energy, which is typically 20\% of the initial proton energy. Hence, the parent cosmic rays have energies of $20-30$~PeV, above the CR {\it knee} at $3-4$~PeV and close to the CR second {\it knee} (or iron {\it knee}) around $100$~PeV~\cite{Apel:2011mi,Apel:2013dga}. Cosmic rays below $100$~PeV are thought to be still dominated by a Galactic population of sources, but this does not rule out a possible sub-dominant extragalactic contribution producing PeV neutrinos inside sources~\cite{Murase:2008yt} or outside sources~\cite{Kalashev:2013vba}. Also, whether the sources are Galactic or extragalactic, $\gamma$-rays should be produced as well as neutrinos. In Refs.~\cite{Prodanovic:2006bq,Gupta:2013xfa,Murase:2013rfa,Anchordoqui:2013lna} it was already pointed out that the ``multi-messenger connection'' between neutrinos and $\gamma$-rays provides important ways to identify or constrain candidate sources of neutrinos.

In the following we will discuss general constraints on the Galactic origin of the IceCube excess from diffuse limits of TeV-PeV $\gamma$-ray observatories. If the observed background neutrino flux is nearly isotropic and Galactic, the Galactic origin is already disfavored by PeV $\gamma$-ray limits. However, we stress that these diffuse $\gamma$-ray constraints are biased in the Northern Hemisphere while most of the 28 IceCube events are located in the Southern Hemisphere. We then discuss a possible association of (part of) the IceCube excess with the (quasi-)diffuse emission from the Galactic Plane (GP) and with two extended GeV $\gamma$-ray emission regions close to the GC, known as the {\it Fermi Bubbles} (FBs). Although there is no statistically significant neutrino event clustering at present~\cite{Aartsen:2013pza}, we show that diffuse TeV to PeV $\gamma$-ray surveys are an important probe to break the degeneracy between Galactic and extragalactic contributions.

In the following we work in Heaviside-Lorentz units and make use of the abbreviation $A_x = A/(10^xu)$, where $u$ is the (canonical) unit of the quantity $A$. 

\section{Neutrino and Gamma-Ray Relation}

In general, $pp$ and $p\gamma$ interactions produce $\gamma$-rays as well as neutrinos. The relative flux of neutrinos and pionic $\gamma$-rays depend on the ratio of charged to neutral pions, $K=N_{\pi^\pm}/N_{\pi^0}$. The relative neutrino flux per flavor depend on the initial mix of $\pi^+$ to $\pi^-$. In the case of $p\gamma$ interactions $\pi^+$ and $\pi^0$ mesons are produced at about the same rate (including direct pion production). Hence $K\simeq1$ and after oscillation the flavor ratio $\nu_e:\bar\nu_e:\nu_\mu:\bar\nu_\mu:\nu_\tau:\bar\nu_\tau$ is approximately $14:4:11:7:11:7$ (in the ``tri-bi-maximal'' approximation). In the case of $pp$ interactions we have $K\simeq2$ and the flavor ratio is equal. In this work, we focus on $pp$ interactions that are thought to be the main hadronic process for Galactic sources detected by $\gamma$-ray observations.  

The pionic $\gamma$-ray production rate can thus be estimated in the following way. In the decay $\pi^0\to\gamma\gamma$ the $\gamma$-ray takes half of the pion energy. As mentioned earlier, each of the three neutrinos from charged pion decay carries about one quarter of the pion's energy. Hence, the relative differential fluxes $J$ of $\gamma$-rays and neutrinos at energies $E_\gamma \simeq 2E_\nu$ are related as
\begin{equation}\label{eq:Jrel}
E_\gamma J_\gamma(E_\gamma) \simeq e^{-\frac{d}{\lambda_{\gamma\gamma}}}\frac{2}{K}\frac{1}{3}\sum_{\nu_\alpha}E_\nu J_{\nu_\alpha}(E_\nu)\,,
\end{equation}
where $d$ is the distance to the source. Here, we account for the absorption of TeV-PeV $\gamma$-rays in radiation backgrounds with interaction length $\lambda_{\gamma\gamma}(E_\gamma)$. We assume that sources are optically thin, which is true for many Galactic CR sources, and for Galactic distances the most important absorption process is the scattering of PeV $\gamma$-rays off the cosmic microwave background with interaction length of about 10~kpc. For sources in or close to the GC, there are additional backgrounds from interstellar radiation fields~\cite{Moskalenko:2005ng,Porter:2008ve}. We do not have to take into account electromagnetic cascades since the reprocessed component appears in the lower-energy range.  Note that the connection between neutrinos and $\gamma$-rays is important even for extragalactic CR sources, and it is especially critical to test the hadronuclear origin of extragalactic neutrino sources~\cite{Murase:2013rfa}.

%%%
\begin{table}[t]
\begin{tabular}{l|cccc}
\hline\hline
Observatory&Position&$[\delta_{\rm min},\delta_{\rm max}]$&Zenith&Refs.\\
\hline
KASCADE&$49.0^\circ$N, $8.4^\circ$E&$[14^\circ,84^\circ]$&$<35^\circ$&\cite{Schatz:2003aw}\\
EAS-TOP&$42.5^\circ$N, $13.5^\circ$E&$[7^\circ,78^\circ]$&$<35^\circ$&\cite{Aglietta:1992,Aglietta:1996ds} \\
GAMMA&$40.5^\circ$N, $44.2^\circ$E&$[10^\circ,71^\circ]$&$<30^\circ$&\cite{Martirosov:2009ni} \\
UMC&$40.2^\circ$N, $112.8^\circ$W&$[0^\circ,80^\circ]$&$<40^\circ$&\cite{Matthews:1990zp}\\
CASA-MIA&$40.2^\circ$N, $112.8^\circ$W&$[-20^\circ,90^\circ]$&$<60^\circ$&\cite{Chantell:1997gs,Borione:1997fy} \\
Milagro&$35.9^\circ$N, $106.7^\circ$W&$[-14^\circ,86^\circ]$&$<50^\circ$&\cite{Atkins:2005wu,Abdo:2008if}\\
Tibet&$30.1^\circ$N, $90.5^\circ$E&$[-20^\circ,80^\circ]$&$<50^\circ$&\cite{Amenomori:2006vd}\\
HEGRA&$28.7^\circ$N, $17.9^\circ$W&$[-6^\circ,64^\circ]$&$<35^\circ$&\cite{Karle:1995,Horns:1999rb,Aharonian:2001ft}\\
GRAPES-3&$11.4^\circ$N, $76.7^\circ$E&$[-14^\circ,36^\circ]$&$<25^\circ$&\cite{Minamino:2009ds}\\
IceCube&South Pole&$[-90^\circ,-60^\circ]$&$<30^\circ$&\cite{Aartsen:2012gka}\\
\hline\hline
\end{tabular}
\caption[]{List of experiments that constrain the isotropic and/or Galactic diffuse TeV-PeV $\gamma$-ray emission shown in Figs.~\ref{fig2} and \ref{fig4}-\ref{fig6}. We order the observatories from North to South and indicate the observed declination range and zenith angle cut used in the corresponding analysis.}\label{tab1}
\end{table}
%%%

The most sensitive experimental technique for the observation of very high energy ($E_\gamma\gg$~TeV) diffuse~\footnote{In the GeV range covered by Fermi, the diffuse $\gamma$-ray background flux means the unresolved flux from the sky, while the diffuse $\gamma$-ray and neutrino backgrounds in the PeV range, which we discuss here, include contributions from resolved and unresolved sources.} or quasi-diffuse $\gamma$-ray fluxes is the detection of extended air showers (EAS) via ground-based detectors (see Ref.~\cite{Aharonian:2008zza} for a review). As a very high energy $\gamma$-ray enters the atmosphere it initiates a cascade via repeated pair-production and bremsstrahlung in the Coulomb field of the nuclei. If the electro-magnetic component reaches the ground level it has spread out to several 100 meters from the shower core and can be detected by a surface array of wire chambers, scintillation detectors and/or water Cherenkov detectors. These $\gamma$-ray showers can be discriminated from the large background of CRs via a simultaneous detection of muons that originate in the muon-rich CR showers. This observational method provides a large field of view (FoV) ($\Omega_{\rm FoV}\sim \mathcal{O}({\rm sr})$) with a high duty cycle ($>90$\%). 

\begin{figure*}[t]\centering
\includegraphics[width=0.48\linewidth]{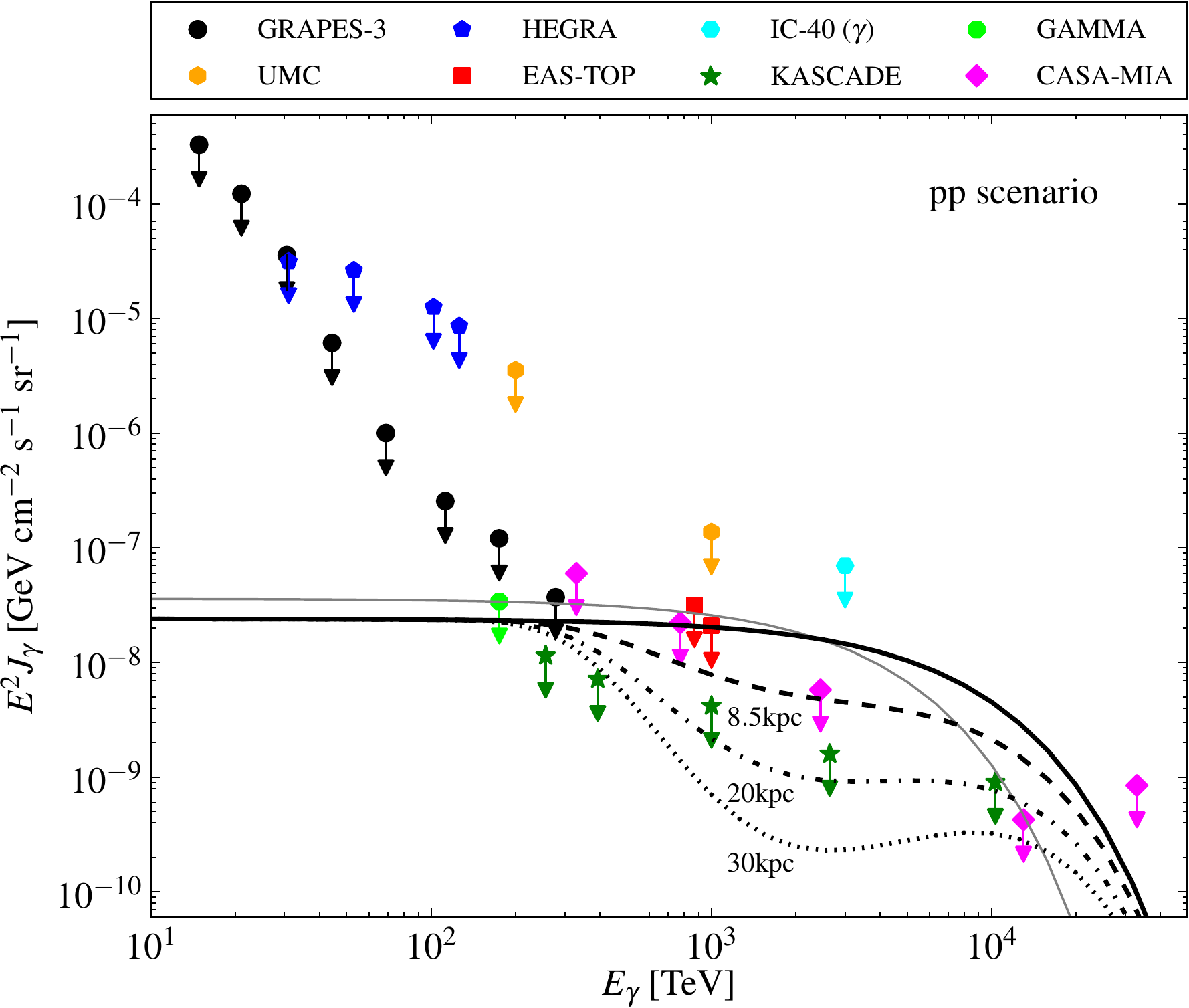}
\hfill
\includegraphics[width=0.48\linewidth]{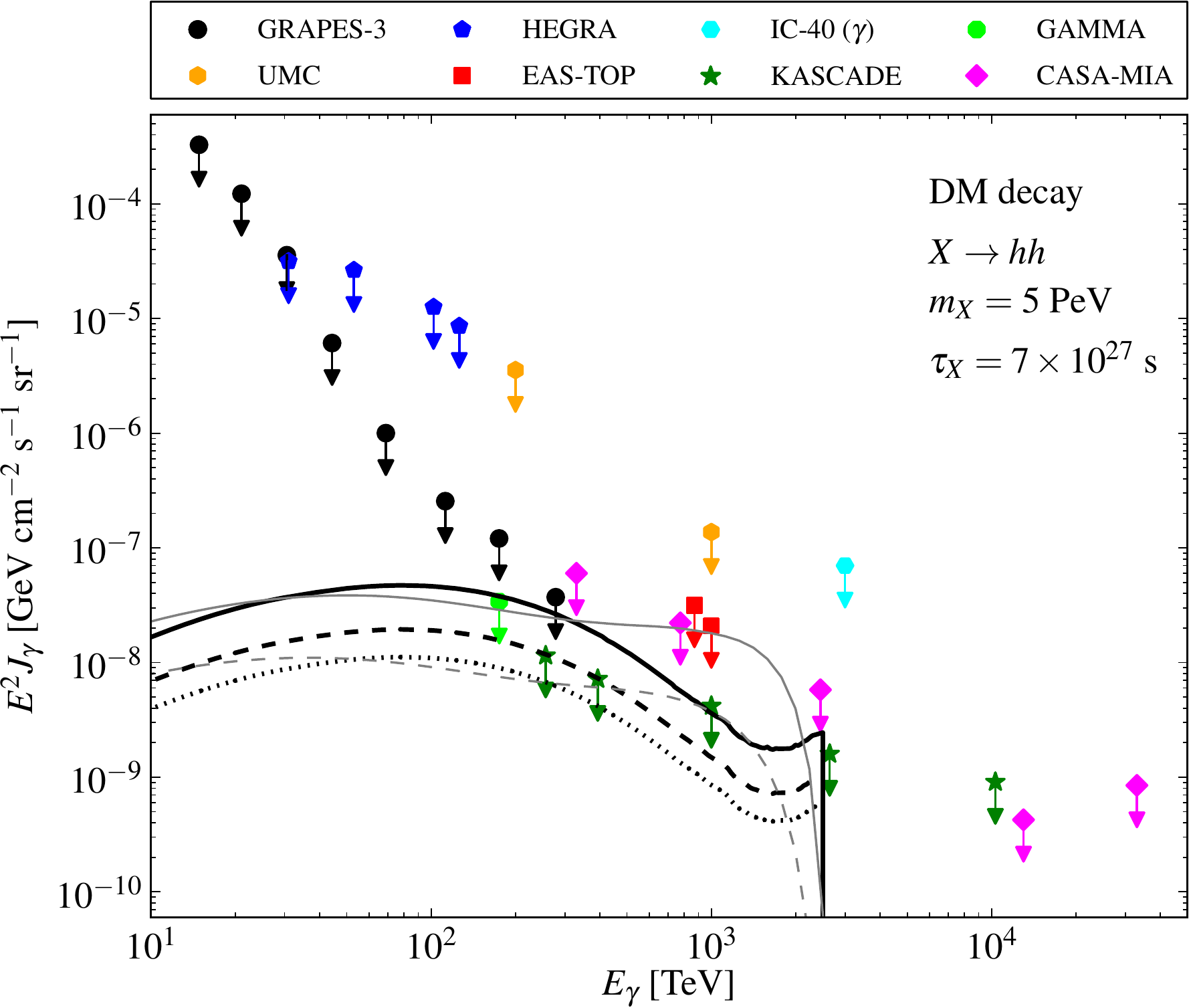}
\caption[]{Measurements of the isotropic diffuse $\gamma$-ray flux in the TeV-PeV range by various experiments (see Table~\ref{tab1}). {\bf Left panel:} The black lines shows the $\gamma$-ray flux corresponding to IceCube's best fit assuming $pp$-interactions ($K=2$) and an exponential cutoff at 6~PeV ({\it i.e.}~3~PeV for neutrinos). We show the unattenuated flux and the flux from $8.5$~kpc, $20$~kpc and $30$~kpc, respectively, taking into account pair production via scattering off CMB photons. For the conversion of photon fractions into photon flux we use the CR flux of Ref.~\cite{Gaisser:2012zz}. For comparison we also show the total neutrino flux as a thin gray line. {\bf Right panel:} Comparison to the Galactic $\gamma$-ray emission of a generic DM decay scenario assuming a scalar $X$ with mass $m_X = 5$~PeV and lifetime $\tau_{X} = 7\times10^{27}$~s. The solid, dashed and dotted black lines show the diffuse emission from the three sky regions divided by the red dashed circles in Fig.~\ref{fig3}. The solid gray line shows the total average neutrino flux, which also accounts for the extragalactic contribution shown separately as a dashed gray line.}\label{fig2}
\end{figure*}

Table~\ref{tab1} gives a list of observatories that provide limits on the diffuse TeV-PeV $\gamma$-ray flux. The maximum (minimum) declination visible by the observatories at latitude $\phi$ and maximum zenith angle $\theta$ is given by $\delta_{\rm max} = \min(90^\circ,\phi+\theta)$ ($\delta_{\rm min} = \max(-90^\circ,\phi-\theta)$). The size of the FoV is given as $\Omega_{\rm FoV} = 2\pi(\sin\delta_{\rm max}-\sin\delta_{\rm min})$. In Fig.~\ref{fig2} we show a summary of upper limits on the isotropic and Galactic diffuse TeV to PeV $\gamma$-ray flux. 

Note that all but one of the observatories in Table~\ref{tab1} are located in the Northern Hemisphere and hence the combined limits on the diffuse $\gamma$-ray flux are biased. The shaded area in Fig.~\ref{fig1} indicates the part of the sky which is presently not constrained by these $\gamma$-ray observatories. The northern edge of the shaded area corresponds to the FoV of CASA-MIA~\cite{Chantell:1997gs}, whereas the southern edge corresponds to the FoV in the search for $\gamma$-rays with IceCube and its surface air shower detector IceTop with the 40 string configuration (IC-40)~\cite{Aartsen:2012gka}. Interestingly, more than half of IceCube's event lie within this ``blind spot''. We would also like to point that the position of one PeV event (20) falls within 10 degree of the hottest spot (with low significance) of IceCube's search for PeV $\gamma$-ray point sources indicated as the red star in Fig.~\ref{fig1}. 

%%%
%%%

\subsection{Quasi-Isotropic Galactic Emission}

The IceCube excess is consistent with an isotropic distribution of arrival directions.  If it is truly isotropic, it is natural to assume that the neutrinos come from extragalactic sources. In principle, however, one could consider possibilities of Galactic sources such as Galactic halos including termination shocks of galactic winds, high-latitude old pulsars, local molecular clouds around the solar system and hot circumgalactic gas.  But, among them, no plausible scenario has been proposed. PeV $\gamma$-ray constraints can strongly support this directly.  

As an astrophysical scenario we briefly discuss the expected neutrino and $\gamma$-ray emission from the Galactic halo following Ref.~\cite{Feldmann:2012rx}. We assume that the ejecta of Galactic supernovae (SN) accelerate CRs to an energy above the CR knee sufficient for the production of PeV neutrinos. (We will provide a more detailed discussion of the maximum CR energy in supernova remnant (SNR) shocks in the following section.) The total CR energy per SN is assumed to be a significant energy fraction $\epsilon_p$ of the initial SN ejecta energy of ${\mathcal E}_{\rm ej}={10}^{51}~{\rm erg}~{\mathcal E}_{\rm ej,51}$. In the following we approximate the source CR spectrum as a power-law normalized as $E_p^2N_p(E_p) \simeq \epsilon_p{\mathcal E}_{\rm ej}(E_p/E_{p,\rm min})^{2-\Gamma}/\mathcal{R}_0$, where we assume that $E_{p,{\rm min}}\sim m_p$ and introduce a bolometric correction factor $\mathcal{R}_0 = (1-{(E_{p, \rm max}/E_{p, \rm min})}^{2-\Gamma})/(\Gamma-2)$ (or $\mathcal{R}_0 = \ln(E_{p,{\rm max}}/E_{p,{\rm min}})$ for $\Gamma=2$).

We now assume that CRs injected over a time scale of $t_{\rm inj}\sim10$~Gyr can be trapped in the Galactic halo~\footnote{Galactic magnetic fields are expected to be weaker at the outer halo with less efficient diffusion. It is not easy for $\sim10$~PeV CRs to be confined for $\sim10$~Gyr, unless the diffusion coefficient in the halo is smaller than that in the disk.} with a gas density $n_{\rm halo}\simeq{10}^{-4.2}~{\rm cm}^{-3}~{(r/R_{\rm vir})}^{-0.8}$~\cite{Werk:2014fza} up to the virial radius $R_{\rm vir}\simeq260$~kpc~\cite{BoylanKolchin:2010ck}. Assuming the present supernova rate of $R_{\rm SN}\sim0.03~{\rm yr}^{-1}$ and its past enhancement $f_{\rm past}\sim3$ the total number of SNRs contributing to the halo emission is $N_{\rm SNR} \simeq f_{\rm past}R_{\rm SN}t_{\rm inj}$. The present energy density of CRs in the halo is thus approximately $N_{\rm SNR}\epsilon_p\mathcal{E}_{\rm ej}/V_{\rm halo}$ with halo volume $V_{\rm halo}\simeq(4\pi/3)R_{\rm vir}^3$. The per flavor and per SNR neutrino spectral emissivity is then~\citep[{c.f.}][]{Murase:2013rfa} $E_\nu^2Q_{\nu_\alpha}\simeq(1/6)\kappa_pc\sigma_{pp}n_{\rm halo}E_p^2N_p(E_p)$, where $E_\nu \simeq 0.05E_p$ and for $pp$ interactions we used the pion ratio $K\simeq2$, mean inelasticity $\kappa_p\simeq0.5$ and cross section $\sigma_{pp}\simeq 3\times10^{-26}~{\rm cm}^2$ around $1$~GeV, increasing to $\sigma_{pp}\simeq 6\times10^{-26}~{\rm cm}^2$ around $E_{\rm kn}$~\cite{Kelner:2006tc}.
The diffuse neutrino spectrum can then be approximated as
\begin{multline}
E_\nu^2 J^{\rm halo}_{\nu_\alpha} \simeq
\frac{N_{\rm SNR}}{4\pi V_{\rm halo}}\int_0^{R_{\rm vir}} {\rm d}r E_\nu^2Q_{\nu_\alpha}\\
\simeq  2.4\times10^{-9}~{\rm GeV}~{\rm cm}^{-2}~{\rm s}^{-1}~{\rm
sr}^{-1}~\epsilon_{p,-1} {\mathcal E}_{\rm ej,51}\\
\times\left(\frac{R_{\rm vir}}{260~{\rm kpc}}\right)^{-2}\left(\frac{f_{\rm past}}{3}\right)\left(\frac{R_{\rm
SN}}{0.03~{\rm yr}^{-1}}\right)\left(\frac{t_{\rm inj}}{10~{\rm Gyr}}\right)\,,
\end{multline}
for $\Gamma=2$, $E_{p,\rm min}\sim m_p$ and $E_{p,\rm max}\sim12$~PeV.

Note that the previous estimate is consistent with results obtained by Ref.~\cite{Feldmann:2012rx} if we adopt $\Gamma=2.4$, but the contribution to the diffuse PeV neutrino flux is negligible for such steeper indices. Thus, it is difficult to explain the IceCube flux, unless the stronger enhancement $f_{\rm past}\sim15$ and $\Gamma=2$ are achieved.  Note that $\sim63$\% of the contributions come from $r<0.1R_{\rm vir}\sim26$~kpc and this fraction is higher when the realistic CR density gradient is taken into account, so PeV $\gamma$-ray observations are important to test this possibility.  Furthermore, as discussed below, the diffuse isotropic $\gamma$-ray background measured by Fermi-LAT can already put strong constraints on this scenario.

In the left panel of Fig.~\ref{fig2} we show the expected isotropic diffuse flux of $\gamma$-rays using the flux~(\ref{eq:ICnu}) and relation~(\ref{eq:Jrel}) for a hypothetical Galactic source distribution. We indicate the absorption effects from pair production in the cosmic microwave background (CMB) via a fiducial distance of $8.5$~kpc (distance of the GC), $20$~kpc and $30$~kpc of the CR interaction site. We also show upper limits on the isotropic diffuse $\gamma$-ray flux by observatories listed in Tab.~\ref{tab1}. Clearly, at PeV energies the required $\gamma$-ray flux is already disfavored by upper limits obtained with two independent measurements, CASA-MIA and KASCADE. Along with the fact that many events come from high latitudes far from the GP, this supports the interpretation of the IceCube excess as an extragalactic flux.

Studies of the isotropic diffuse high-energy $\gamma$-ray emission in the near future with the HAWC observatory~\cite{DeYoung:2012mj} located in Mexico ($19.0^\circ$N, $97.3^\circ$W) and the far future with the air shower array LHAASO~\cite{Zha:2012wp} in China (Yunnan Province, $27.8^\circ$N, $99.7^\circ$E) or HiSCORE~\cite{Tluczykont:2012nm} can greatly improve the present limits. It is also important for the air shower arrays to cover the sky region where neutrinos were found by IceCube. In the Southern Hemisphere the full IceCube detector (IC-86) has a higher sensitivity and larger FoV (zenith angle range $\theta<45^\circ$) available for the study of diffuse PeV $\gamma$-ray emission~\cite{Aartsen:2012gka}. The combined search by HAWC (assuming a zenith angle range $\theta<50^\circ$) and IC-86 will further reduce the ``un-charted'' high-energy $\gamma$-ray sky as indicated by the reduced dark-shaded area in Fig.~\ref{fig1}. Note that a possible location of HiSCORE in Malarg\"ue, Argentina ($35.5^\circ$S, $69.6^\circ$W) with a zenith angle range of $30^\circ$ would fully cover this $\gamma$-ray ``blind spot''.

If the IceCube excess has a hadronuclear ($pp$) origin it is even possible to constrain this scenario via the diffuse isotropic gamma-ray background measured by Fermi-LAT~\cite{Abdo:2010nz}. The secondary $\gamma$-ray and neutrino spectra from $pp$ collisions follow the initial CR spectrum $\propto E^{-\Gamma}$ with $\Gamma \gtrsim 2$. Hence, normalizing the neutrino spectrum to the IceCube excess in the TeV-PeV range fixes the spectra also in the sub-TeV range. In fact, the Galactic $pp$ origin of the IceCube excess can be consistent with the preliminary Fermi data in the $(0.1-1)$~TeV range~\cite{Murase:2013rfa} only for hard CR power-law spectra, $\Gamma\gtrsim 2$. This scenario can be excluded via future constraints on $\Gamma$ with continued neutrino observation in the sub-PeV range and by limiting the contribution of candidate neutrino sources to the isotropic gamma-ray background.

%%%
\begin{figure*}[t]
\includegraphics[width=0.8\linewidth]{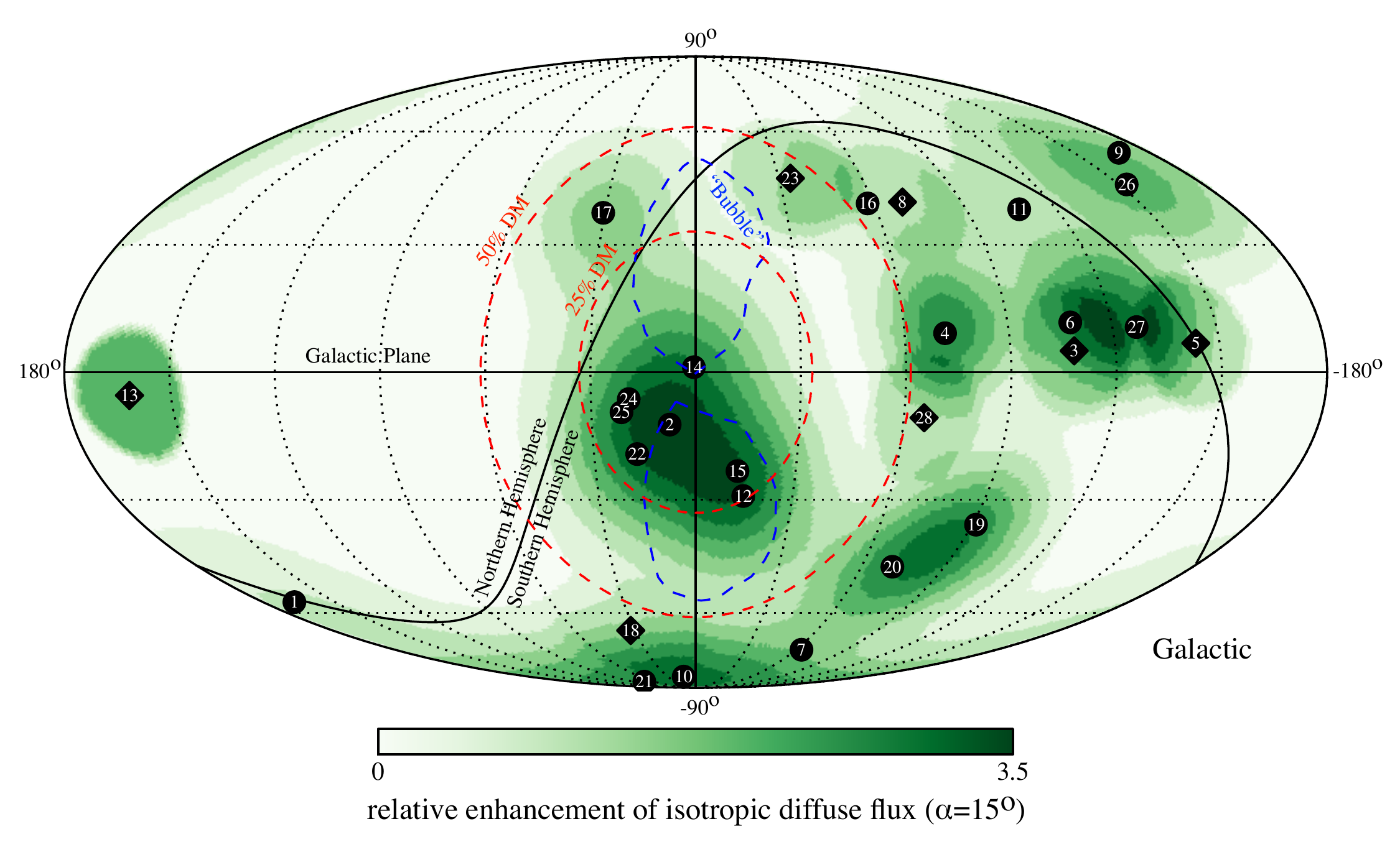}
\caption[]{The fluctuation of the non-isotropic diffuse flux described by Eq.~(\ref{eq:mod}) assuming extended emission regions with radius $15^\circ$ around each direction in the sky. The events are weighted according to the approximation described in the text. The blue dashed lines indicate the position of the FBs. The red dashed lines show the GC region containing 25\% and 50\% of the emission from DM decay in the Galactic halo.}\label{fig3}
\end{figure*}
%%%

Another possible Galactic source of the IceCube excess consists of very heavy dark matter (DM) in the Galactic halo, which decay or annihilate into Standard Model particles~\cite[{\it e.g.},][and references therein]{Murase:2012xs}. Depending on the particular model, their particle properties can be probed by neutrino and $\gamma$-ray observations. The emission will be very extended and can be compared to the limits on the isotropic diffuse $\gamma$-ray emission. In Fig.~\ref{fig3} we indicate the Galactic Center region containing 25\% and 50\% of the local DM decay from the Galactic halo. The two-body decay of the DM particle may produce PeV neutrino line features with some continua~\cite{Feldstein:2013kka,Esmaili:2013gha,Bai:2013nga}. For instance, PeV DM gravitinos in $\mathcal{R}$-parity violating supersymmetric models would decay into neutrinos and/or photons. Note that this would also result in high-energy $\gamma$-rays that may include a PeV $\gamma$-ray line feature~\cite{Buchmuller:2007ui}.

In the following we will discuss a simple DM scenario consisting of a scalar particle $X$ with mass $m_X = 5$~PeV and lifetime $\tau=7\times10^{27}$~s that decays into two Standard Model Higgs $h$~\cite{Bai:2013nga}. This scenario produces a flat secondary flux of neutrinos with $E_\nu<m_X/2$ that can resemble the spectral features of the IceCube excess. We determine the energy distributions $Q_\nu(E_\nu)$ and $Q_\gamma(E_\gamma)$ of secondary neutrinos and $\gamma$-rays, respectively, via the Monte Carlo code PYTHIA~\cite{Sjostrand:2007gs}. The $4\pi$-averaged diffuse Galactic emission can then be calculated as
\begin{equation}\label{eq:DMgal}
J^{\rm gal}_{\nu/\gamma}(E) = \frac{Q_{\nu/\gamma}(E)}{8\pi m_X\tau_X}\int\limits_0^\infty{\rm d}s\int\limits_{-1}^1{\rm d}c_\alpha\, \rho(r(s,c_\alpha))\,,
\end{equation}
where $\rho(r)$ is the spherical mass density of the Galactic DM halo at radius $r$, which can be parametrized by the line-of-sight distance $s$ and angular distance $\alpha$ towards the GC as $r^2(s,\cos\alpha) =s^2+R_\odot^2-2sR_\odot\cos\alpha$. We use the Einasto profile~\cite{Graham:2006ae} $\rho(r)\propto\exp[-(2/\beta)(r/20{\rm kpc})^\beta]$ with $\beta=0.17$ and normalization $\rho(R_\odot)=0.4{\rm GeV}/{\rm cm}^3$.

For the correct normalization of the neutrino emission it is also necessary to include extragalactic contributions which are simply given by
\begin{equation}\label{eq:DMxgal}
J^{\rm xgal}_{\nu}(E)\nu) = \frac{\Omega_{\rm DM}\rho_{\rm cr}}{4\pi m_X\tau_X}\int\limits_0^\infty\frac{{\rm d}z}{H(z)}Q_\nu((1+z)E_\nu)\,,
\end{equation}
where $H^2(z) = H^2_0[\Omega_\Lambda+(1+z)^3\Omega_{\rm m}]$ is the Hubble constant with $\Omega_{\rm m}\simeq 0.3$, $\Omega_\Lambda\simeq 0.7$ and $H_0 \simeq 70\, {\rm km}\,{\rm s}^{-1} {\rm Mpc}^{-1}$. The comoving DM density is parametrized via the critical density $\rho_{\rm cr}\simeq5\times10^{-6} {\rm GeV}/{\rm cm}^3$ and DM fraction $\Omega_{\rm DM}\simeq 0.27$~\cite{Beringer:1900zz}. Note, that the extragalactic contributions in the form of $\gamma$-rays (and electrons/positrons) will not directly be observable, but initiate electro-magnetic cascades in the cosmic radiation backgrounds. This will populate the extragalactic $\gamma$-ray background in the GeV-TeV energy range. The extragalactic $\gamma$-ray background inferred by Fermi-LAT can thus also constrain this scenario~\cite{Murase:2012xs}.

In the right panel of Fig.~\ref{fig2} we show the total neutrino flus as a sum of Eqs.(\ref{eq:DMgal}) and (\ref{eq:DMxgal}) indicated as a solid gray line. For comparison, the extragalactic contribution is indicated separately as a dashed gray line. The solid, dashed and dotted black lines show the diffuse $\gamma$-ray emission from the three sky regions divided by the red dashed circles in Fig.~\ref{fig3}. This indicates the increased diffuse emission towards the GC. Note, that the GC itself is only barely visible by the experiments listed in the figure. This scenario is hence marginally consistent with the non-observation of PeV $\gamma$-rays. However, an observatory in the Southern Hemisphere covering the GC with a $0.1-1$~PeV $\gamma$-ray sensitivity comparable to that of the KASCADE array would be sufficient to constrain this DM model. Moreover, the all-sky averaged PeV $\gamma$-ray flux from DM decay is in reach of future observatories like HiSCORE or LHAASO.

Note that, in this specific DM decay scenario, the total neutrino flux is a factor of two higher than the generated $\gamma$-ray flux since the neutrino flux includes extragalactic contributions. Although we only consider $X \to hh$ for demonstration, different DM scenarios with line features or extended decay channels, {\it e.g.}~$X \to\tau^+\tau^-$ can lead to increased PeV $\gamma$-ray emission that can already be excluded by diffuse TeV-PeV $\gamma$-ray limits.

\subsection{Non-Isotropic Galactic Emission}

In the previous section, we demonstrated the power of PeV $\gamma$-ray searches. If the observed neutrino emission is largely isotropic and Galactic, it contradicts existing PeV $\gamma$-ray measurements, supporting extragalactic scenarios.  In principle, the observed events could come from Galactic sources that do not accidentally exist in the sky region covered by various air shower arrays. Indeed, more than half of IceCube's events lie within this ``blind spot'', so that we cannot rule out such a possibility. But, since many events appear significantly out of the GP, powerful Galactic accelerators seem to be needed even at high latitude, which is theoretically challenging. PeV $\gamma$-ray observations covering the IceCube sky should enable us to support or exclude such speculations.

A more natural situation may be that the observed IceCube excess consist of a superposition of Galactic and extragalactic events. In fact, IceCube observes a slight excess of events in the Southern Hemisphere and a weak clustering of cascades close to the GC. However, we caution that there are no statistically significant fluctuations at present and accumulation of further neutrino data is required. Nevertheless, it is interesting to see how possible event clustering by Galactic sources are tested by combing PeV neutrino and $\gamma$-ray observations. Also, event clustering due to Galactic sources are theoretically possible, especially for the GP and the FBs. Hereafter we consider how we identify potential Galactic neutrino sources, which will become important to break a degeneracy between extragalactic and Galactic contributions.  

In general, event clustering can originate from nearby bright sources or a spatial clustering of sources with average brightness. In the following we consider the possibility that the neutrino sky is a superposition of event clusters of Galactic sources (point-like or extended) and an isotropic extragalactic background. As a result, certain regions in the IceCube sky might correspond to a higher non-isotropic diffuse limit than the all-sky average. To account for local fluctuations we first construct weights of the 28 events that approximate the probability that the event is signal-like. From the diffuse flux and background expectation in the histograms of Fig.~6 in Ref.~\cite{Aartsen:2013pza} we extract the probabilities $P_1(E_{\rm dep})$ and $P_2(\delta)$ that an event with deposited energy $E_{\rm dep}$ and declination band $\delta$ is part of the excess. For this we assume that the atmospheric neutrino background (mostly $\nu_\mu$), atmospheric muon background and the expected diffuse flux splits between tracks and cascade as $2/3:1/3$, $1:0$ and $2/9:7/9$, respectively. We then assign each event a weight $w_i \propto (P_1+P_2)/2$, which we normalize such that $\sum_i w_i = (28-10.6)/28$, reproducing the expected overall background of events. Note that this approximation can be easily extended to the true signal probability distribution in $E_{\rm dep}$ and $\sin\delta$ that can be derived by Monte-Carlo simulations. With this method the weights of muons are much lower than the weights of the cascades except for event 13 that is an up-going muon close to the GP. This upward going muon has a large relative weight due to the small background from atmospheric muons that can not penetrate the Earth far below the horizon.

We then derive non-isotropic diffuse limits in the following way. We calculate the number of signal events $n_{\Delta\Omega}$ in an area $\Delta\Omega$ as
\begin{equation}\label{eq:n}
n_{\Delta\Omega} = \sum_{i=1}^{28}\int\limits_{\Delta\Omega}{\rm d}\Omega w_if_i(\Omega)\,,
\end{equation}
where the sum runs over all 28 events with weight $w_i$ and angular probability distribution $f_i$ taken into account the angular reconstruction uncertainty. We approximate the $f_i$ in the calculation by (normalized) Gaussian distributions. The quasi-diffuse flux (including the isotropic component) in this region $\Delta\Omega$ is then approximated from the IceCube excess~(\ref{eq:ICnu}) as
\begin{equation}\label{eq:mod}
\frac{J^{\Delta\Omega}_\alpha(E_\nu)}{J^{\rm IC}_{\alpha}(E_\nu)} \simeq \frac{n_{\Delta\Omega}}{n_{4\pi}}\frac{4\pi}{\Delta\Omega}\,.
\end{equation}
With this we can construct an un-binned scan of non-isotropic diffuse fluctuations over the full sky. Around each point in the sky we define an area with a half opening angle $\alpha$ and size $\Delta\Omega=2\pi(1-\cos\alpha)$. For this area we calculate the fluctuation via Eqs.~(\ref{eq:n}) and (\ref{eq:ICnu}) for a fixed angle $\alpha$. In Figure \ref{fig3} we show the relative local fluctuation of the diffuse flux for a fiducial value of $\alpha=15^\circ$. In some parts of the sky, in particular in the extended region around the GC, the diffuse flux is enhanced by more than a factor $\sim 3$. Also visible is the enhancement via the track event 13. Since the direction of the track is known within about $1.2^\circ$ the procedure of integrating events within $15^\circ$ results in a disc of the same radius centered at the track. Different emission profiles of the extended regions can be easily adopted in this calculation. 

The statistical uncertainty of the IceCube excess of 28 events with $10.6$ expected background events is very high. Hence, the local fluctuations visible in Fig.~\ref{fig1} can be entirely statistical as was also pointed out in Ref.~\cite{Aartsen:2013pza}. However, it is also possible that these fluctuations indicate a weak non-isotropic diffuse contribution. The most prominent local fluctuation in Fig.~\ref{fig3} corresponds to an extended region of seven cascades close to the line of sight of the Galactic Center. Other, but weaker local fluctuation exist across the sky, including an elongated cluster of two cascades and two muons that appears within $15^\circ$ above the Galactic Plane at longitudes $210^\circ\lesssim \ell \lesssim270^\circ$. 

In the following, motivated by theoretical expectations, we will discuss if these fluctuations could be related to a diffuse emission from the GP or the FBs.

\subsubsection{Galactic Plane}

The GP contains a lot of candidate sources of Galactic CR accelerators, including SN remnants (SNRs) and pulsar wind nebulae (PWNe).  Indeed, various Galactic TeV $\gamma$-ray sources have been identified by imaging atmospheric Cherenkov telescopes~\cite{Aharonian:2008zza,Aharonian:2005jn,Aharonian:2005kn}, which may also be TeV-PeV neutrino emitters~\cite{Kistler:2006hp,Kappes:2006fg}. Galactic CRs escaping from accelerators interact with the interstellar gas via hadronuclear interactions during their propagation~\citep[{\it e.g.},][]{Stecker:1978ah,Domokos:1991tt,Berezinsky:1992wr,Ingelman:1996md,Evoli:2007iy}. The average gas density in the gas scale height of $\sim200$~pc is $n\simeq1~{\rm cm}^{-3}$, and CRs confined in the larger scale of $\sim4$~kpc are thought to produce Galactic diffuse emissions in $\gamma$-rays and neutrinos, as observed by Fermi in the GeV range~\cite{FermiLAT:2012aa}. As noted above, at present the IceCube excess shows no significant event clustering along the GP within latitude $|b|<2.5^\circ$~\cite{Aartsen:2013pza}. However, an extension of the GP region to higher declinations of $|b|\lesssim10^\circ$ could account for up to 13 events within uncertainties. 

A guaranteed contribution of a non-isotropic diffuse neutrino emission along the GP are hadronic interactions of CRs during propagation in the interstellar medium~\citep[{\it e.g.},][]{Stecker:1978ah,Domokos:1991tt,Berezinsky:1992wr,Ingelman:1996md,Evoli:2007iy}. The emission profile follows the gas density in the Milky Way, which is centered along the GP. We assume a gas density in the Milky Way within a radius of $R_{\rm MW}\simeq17$~kpc that decreases above the plane as $n(z) \propto \exp(-z/h)$ with $h\simeq 0.1$~kpc. The inelastic $pp$ cross section can be approximated at $\sigma \simeq 34 + 1.88L +0.25L^2$~mb for $L=\ln (E_p/{\rm TeV})$~\cite{Kelner:2006tc}. The flux is expected to be maximal in the direction of the GP and quickly decays to higher latitudes. The minimum is reached for directions orthogonal to the GP with $J^{\rm min}_\nu(E_\nu)\simeq (0.2/25.5)J^{\rm max}_\nu(E_\nu)$.

We approximate the flux of CR nucleons throughout the Galaxy via the locally observed flux that can be parametrized as \mbox{$E_NJ_N\simeq1.03/({\rm cm}^{2}~{\rm s}~{\rm sr})(E_N/{\rm GeV})^{-\gamma}(1+(E/E^*)^3)^{-\delta/3}$} with $\gamma=1.64$, $\delta=0.67$ and $E^*=0.9$~PeV~\cite{Gaisser:2013ira}. The diffuse per-flavor neutrino flux can then be approximated as $E_\nu^2J_\nu \simeq (1/6)\kappa_p\langle\tau\rangle E_N^2J_N$ where $E_\nu \simeq 0.05E_N$, $\kappa_p\simeq0.5$, and $\langle\tau\rangle$ is the optical depth averaged over Galactic latitude $|b|<2^\circ$. A numerical integration gives the average optical depth of $\langle\tau\rangle\simeq5.8\times10^{-4}$ (compared to $\langle\tau\rangle\simeq3.2\times10^{-4}$, $2.0\times10^{-4}$ or $ 4.5\times10^{-5}$ averaged over $|b|<5^\circ$, $10^\circ$ or the full sky, respectively). We then arrive at a diffuse GP flux of
\begin{multline}\label{eq:JCR}
E_\nu^2J^{{\rm CR}}_{\nu_\alpha} \simeq 8.8\times10^{-8}{\rm GeV}~{\rm cm}^{-2}~{\rm s}^{-1}~{\rm sr}^{-1}\\
\times\left(\frac{E_\nu}{{\rm TeV}}\right)^{-0.64}\left(1+\left(\frac{E_\nu}{45~{\rm TeV}}\right)^3\right)^{-0.22}\!\!\!\!\!.
\end{multline}
At $E_\nu \simeq 200$~TeV, {\it i.e.}~at neutrino energies corresponding to the CRs at the knee $E_{\rm kn}$ the energy density of secondary neutrinos (per flavor) is $J_{\nu_\alpha}\simeq 0.14J^{\rm IC}_{\nu_\alpha}$. At $E_\nu \simeq 60$~TeV the flux is larger, $J_{\nu_\alpha}\simeq0.65J^{\rm IC}_{\nu_\alpha}$, and hence close to the IceCube excess, but at PeV energies it is only $J_{\nu_\alpha}\simeq0.02J^{\rm IC}_{\nu_\alpha}$. Hence, an interpretation of the local IceCube fluctuation as a diffuse emission from hadronic CR interactions in the GP is unlikely. It would require an increased scale height $h\simeq$1-2~kpc to match the scatter of the IceCube events off the GP and the nominal target density $n_0$ in the plane would have to be larger by an order of magnitude to match the observed flux level at PeV energies. Even then, the steep spectral index with $\Gamma\simeq2.6-3.3$ poses a problem in explaining the data over the TeV to PeV range.

%%%
\begin{figure}[t]
\includegraphics[width=\linewidth]{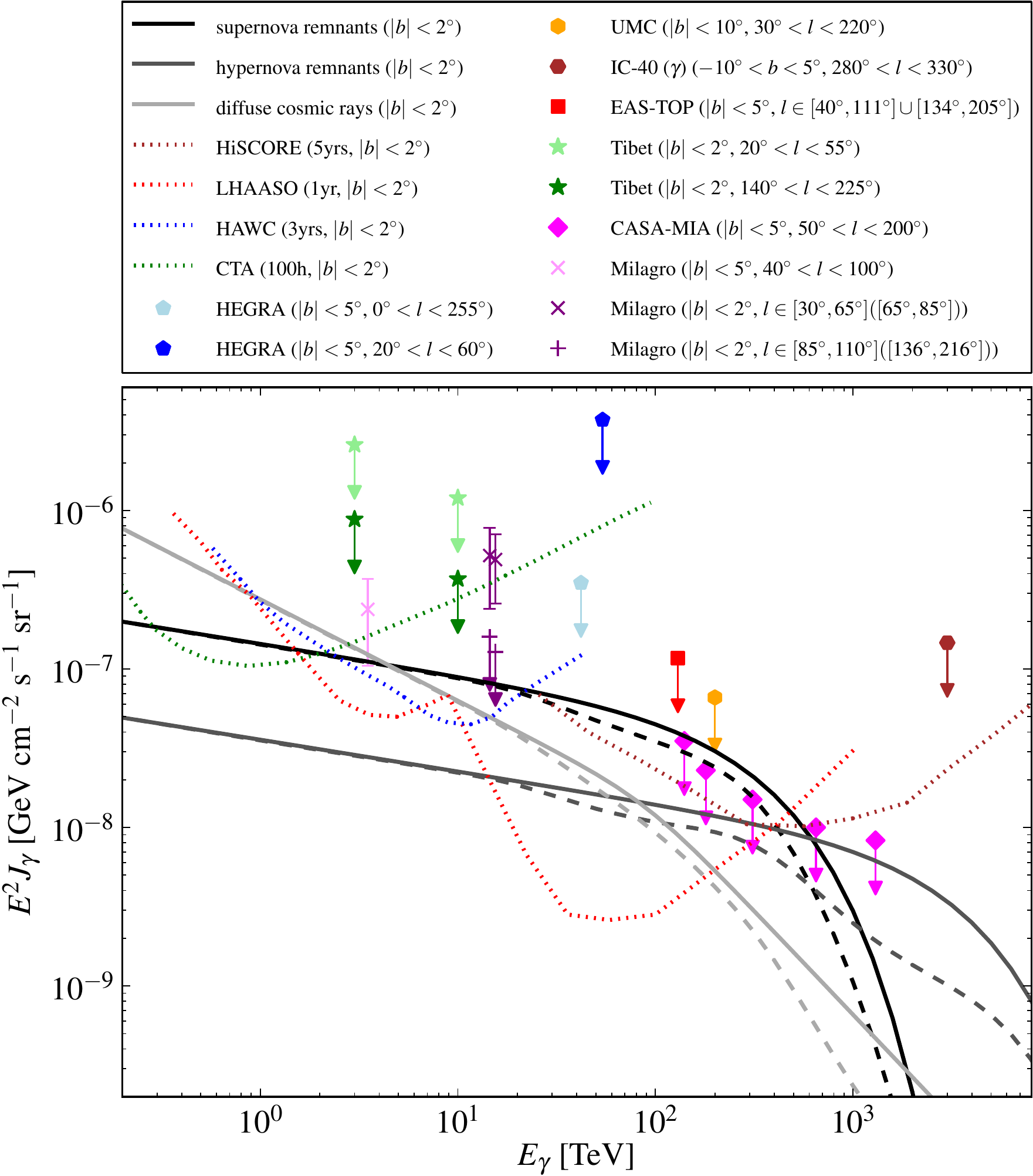}
\caption[]{Diffuse measurements of the $\gamma$-ray flux in the GP in comparison to the expected diffuse flux from the propagation of Galactic CRs (light-gray lines) and from Galactic SNRs (black lines) and HNRs (dark-gray lines) with power index $\Gamma=2.2$. The solid lines indicate the estimate in Eqs.~(\ref{eq:JCR}), (\ref{eq:SNR}) and (\ref{eq:HNR}) using relation (\ref{eq:Jrel}) without attenuation and the dashed lines indicate the contribution from a source at the GC. We adopt the calculation of Ref.~\cite{Moskalenko:2005ng} for the interstellar radiation field on top of the CMB. We also show estimates of the sensitivity of CTA (green dotted), HAWC (blue dotted), LHAASO (red dotted) and HiSCORE (brown dotted) w.r.t.~the diffuse TeV-PeV $\gamma$-ray emission in the GP ($|b|<2^\circ$). Note that the model-dependent theoretical fluxes are averaged over Galactic longitude and latitude $|b|<2^\circ$, whereas the measurements only apply to the intersection of the GP with the FoV and in some case extend to larger absolute latitudes as indicated in the plot ({\it cf.}~Fig.~\ref{fig5}). Extending the GP to $|b|<5^\circ$ or $|b|<10^\circ$ reduces the theoretical fluxes (only $\pi^0$-decay and ignoring absorption) by about a factor 2 or 3, respectively. The relative intensity of the diffuse flux between Galactic Center and anti-Center is less than $\pm25$\% (see text).}\label{fig4}
\end{figure}
%%%

In contrast, unresolved Galactic point sources may give some contributions to the observed IceCube excess.  Note that they may have much harder intrinsic CR spectra. Modeling of the propagation of Galactic CRs predict time-averaged intrinsic CR spectra with $\Gamma\simeq 2.2-2.4$~\cite{Blasi:2011fm}. For example, let us consider SNRs that have been believed to be the main origin of Galactic CRs~\cite{bz34,hay56}. SNRs are distributed in the GP within the radius $R_{\rm MW}$ and height of $|h|\lesssim100$~pc, suggesting that emissions are clustered within $|b|\lesssim2^\circ$~\cite{Evoli:2007iy}. However, bright close-by sources in the local spiral arm of the Milky Way may appear at higher latitudes. In fact, the fluctuation at Galactic longitudes $\ell\simeq240^\circ$ visible in Fig.~\ref{fig3} is close to the direction of the Local Arm. On the other hand, we also point out that there is no apparent neutrino clustering in the opposite direction at $\ell\simeq60^\circ$.

For typical nucleon densities of $n = 1~{\rm cm}^{-3}~n_0$ a significant energy fraction $\epsilon_p$ of the initial SN ejecta energy of ${\mathcal E}_{\rm ej}={10}^{51}~{\rm erg}~{\mathcal E}_{\rm ej,51}$ can have been transferred to CRs by the end of the Sedov phase. Note that the ejecta velocity is $V_{\rm ej}\simeq{10}^{4}~{\rm km}~{\rm s}^{-1}~{\mathcal E}_{\rm ej,51}^{1/2}M_{{\rm ej},\odot}^{-1/2}$ for the mass of the ejecta $M_{\rm ej} = M_{{\rm ej},\odot}M_\odot$. The Sedov radius is $R_{\rm Sed}={(3M_{\rm ej}/4\pi n)}^{1/3}\simeq2.1~{\rm pc}~M_{{\rm ej},\odot}^{1/3}n_0^{-1/3}$ corresponding to the deceleration time of $t_{\rm Sed}\simeq200~{\rm yr}~{\mathcal E}_{\rm ej,51}^{-1/2}M_{{\rm ej},\odot}^{5/6}n_0^{-1/3}$~\cite{tay50,sed46}. The shock velocity $V_s$ decreases as $\propto{(R/R_{\rm Sed})}^{-3/2}$ after $t_{\rm Sed}$. In the Sedov phase, assuming the Bohm limit and a parallel shock, the maximal proton energy is estimated to be $E_{p,{\rm max}}\simeq(3/20)eBRV_{s}/c$~\cite{gai90}, where the magnetic field is parametrized as $B=\sqrt{\varepsilon_Bnm_pV_{s}^2}\simeq0.46~{\rm mG}~\varepsilon_{B,-2}^{1/2}n_0^{1/2}{\mathcal E}_{\rm ej,51}^{1/2}M_{{\rm ej},\odot}^{-1/2}{(R/R_{\rm Sed})}^{-3/2}$ and $\varepsilon_B$ is the fraction of the energy density carried by the magnetic field in the shock. This gives the final estimate of $E_{p,{\rm max}}\simeq4.5~{\rm PeV}~\varepsilon_{B,-2}^{1/2}M_{{\rm ej},\odot}^{-2/3}{\mathcal E}_{\rm ej,51}n_0^{1/6}{(R/R_{\rm Sed})}^{-2}$ which is close to the CR {\it knee}. 

As discussed before, the per flavor neutrino spectral emissivity is given as $E_\nu^2Q_{\nu_\alpha}\simeq(1/6)\kappa_pc\sigma_{pp}nE_p^2N_p(E_p)$. Effective CR acceleration to very high energies ceases at the beginning of the snowplow phase at $t_{\rm sp}\simeq4\times{10}^{4}~{\rm yr}~{\mathcal E}_{\rm ej,51}^{4/17}n_0^{-9/17}$~\cite{Blondin}. For a local SN rate of $R_{\rm SN}\sim0.03~{\rm yr}^{-1}$ the number of active SNRs is of the order of $N_{\rm SNR}\simeq R_{\rm SN}t_{\rm sp}\simeq1200$. The cumulative diffuse flux from SNRs in the GP with $\Delta\Omega_{\rm GP}\simeq0.44$~sr ($|b|<2^\circ$) can then be estimated as
\begin{multline}\label{eq:E2J}
E_\nu^2 J^{\rm SNR}_{\nu_\alpha} \sim \frac{N_{\rm SNR}\langle r_{\rm los}\rangle}{4\pi V_{\rm GP}}E_\nu^2 Q_{\nu_\alpha}\\
\simeq  2.2\times10^{-6}~{\rm GeV}~{\rm cm}^{-2}~{\rm s}^{-1}~{\rm sr}^{-1}\frac{1}{\mathcal{R}_0}\left(\frac{E_{\nu}}{E_{\nu,{\rm min}}}\right)^{2-\Gamma}\\
\times\epsilon_{p,-1} {\mathcal E}_{\rm ej,51}N_{\rm SNR, 3}\langle r_{\rm los}\rangle_1\,,
\end{multline}
with $E_{\nu, {\rm min}} \simeq 0.05E_{p, {\rm min}}$ and $V_{\rm GP} \simeq 2\pi R_{\rm MW}^2 h$. Here we introduce the line-of-sight distance $\langle r_{\rm los}\rangle$ averaged over Galactic longitude and latitude $|b|<2^\circ$~\cite{Casanova:2007cf}. For a homogeneous distribution within radius $R_{\rm MW}\simeq17$~kpc and scale height $h\simeq0.1$~kpc we derive $\langle r_{\rm los}\rangle\simeq 7.5$~kpc (compared to $\langle r_{\rm los}\rangle\simeq 4.0$~kpc or $2.4$~kpc for $|b|<5^\circ$ or $10^\circ$, respectively). Assuming $\Gamma=2.2$, $\mathcal{R}_0\simeq4.8$ and $\langle r_{\rm los}\rangle\simeq 7.5$~kpc we hence have a flux of
\begin{equation}\label{eq:SNR}
E_\nu^2J^{\rm SNR}_{\nu_\alpha} \simeq 2.5\times{10}^{-8}~{\rm GeV}~{\rm cm}^{-2}~{\rm s}^{-1}~{\rm sr}^{-1}\left(\frac{E_{\nu}}{0.1{\rm PeV}}\right)^{-0.2}\!\!\!,
\end{equation}
with exponential cutoff at $E_{\nu,{\rm max}}\simeq0.2$~PeV.

%%%
\begin{figure}[t]
\includegraphics[width=\linewidth]{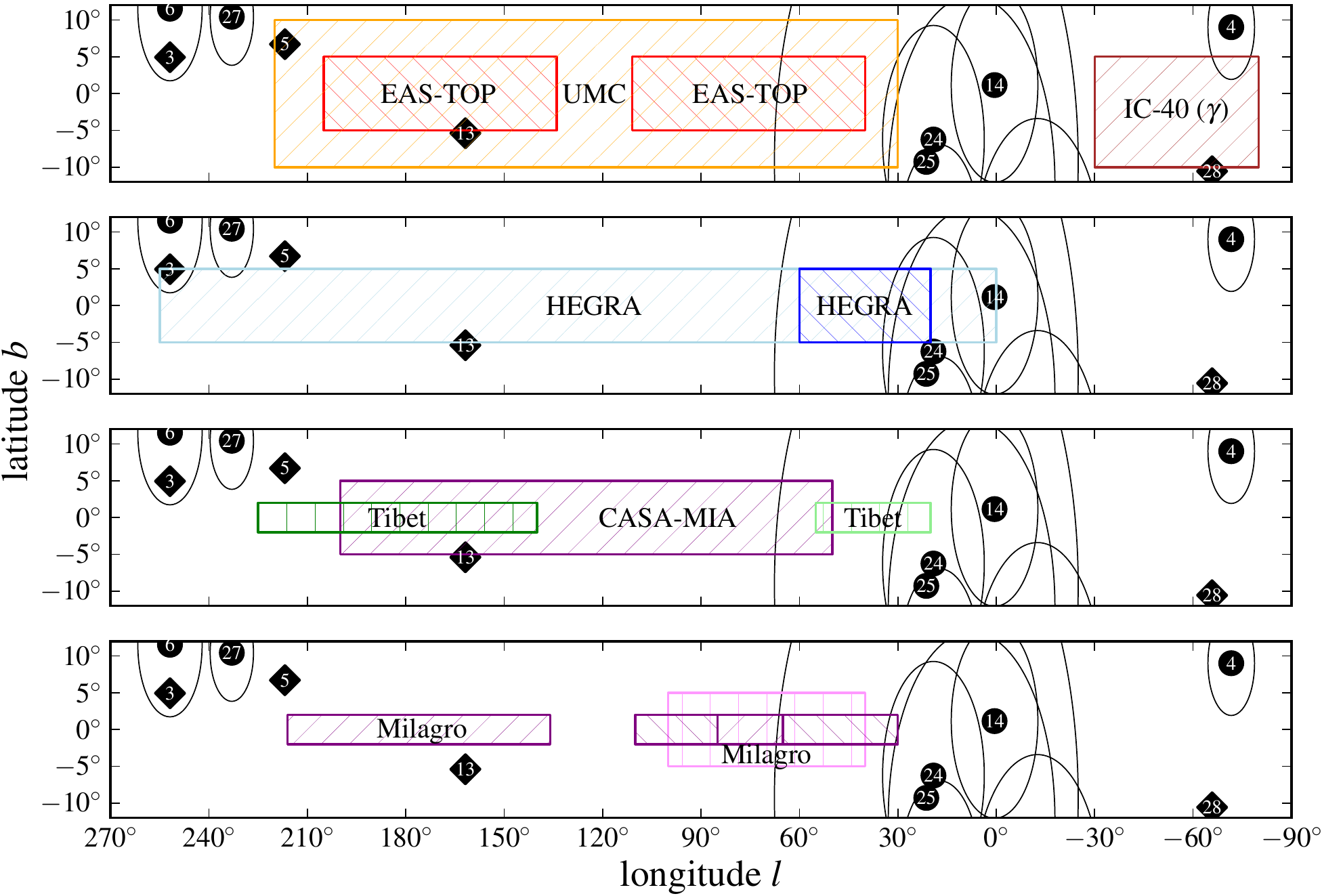}
\caption[]{The on-source regions of GP diffuse emission used for the experimental results shown in Fig.~\ref{fig4} using the same color-coding. We also show the distribution of IceCube events in the vicinity of the GP ({\it cf}.~Fig.~\ref{fig1}). The circled areas indicate the uncertainty of the cascade reconstruction as in Fig.~\ref{fig4}. Note that the limits on diffuse $\gamma$-ray emission along the GP from HEGRA~\cite{Horns:1999rb} assume a larger zenith angle range than for the isotropic diffuse emission listed in Tab.~\ref{tab1}.}\label{fig5}
\end{figure}
%%%

The required CR energy of $20-30$~PeV for the production of 1~PeV neutrinos can be reached by hypernovae (HN) with energies of ${\mathcal E}_{\rm ej}\gtrsim10^{52}$~erg~\cite{Dermer:2000gn,ew01,sve03}. One should keep in mind that most of the HNe are non-relativistic, and trans-relativistic SNe, which have also been suggested as powerful CR accelerators~\cite{Murase:2008mr,Wang:2007ya,Budnik:2007yh}, are much rarer and not necessarily HNe, {\it e.g.}, GRB 060218 with ${\mathcal E}_{\rm ej}\sim2\times{10}^{51}$~erg~\cite{Mazzali:2006tk}.  
It has been suggested that unidentified TeV $\gamma$-ray sources that may include HN remnants (HNRs) may explain a part of the observed neutrino events~\cite{Fox:2013oza}. The HN rate is $\sim1-2$\% of the SN rate~\cite{Guetta:2006gq,Arcavi:2010wd}, so we expect $N_{\rm HNR}\sim20-40$. Taking a fiducial value of $N_{\rm HNR}=30$, a power index $\Gamma=2.2$ and $\langle r_{\rm los}\rangle\simeq 7.5$~kpc we arrive at
\begin{equation}\label{eq:HNR}
E_\nu^2J^{\rm HNR}_{\nu_\alpha} \simeq 6.2\times{10}^{-9}~{\rm GeV}~{\rm cm}^{-2}~{\rm s}^{-1}~{\rm sr}^{-1}\left(\frac{E_{\nu}}{0.1{\rm PeV}}\right)^{-0.2}\!\!\!,
\end{equation}
with exponential cutoff at $E_{\nu,{\rm max}}\simeq2$~PeV.

In Figure~\ref{fig4} we show the associated flux of diffuse Galactic CRs and from SNRs/PWNe and HNRs from Eqs.~(\ref{eq:JCR}), (\ref{eq:HNR}) and (\ref{eq:SNR}) using relation (\ref{eq:Jrel}) in comparison to experimental observations of TeV-PeV $\gamma$-rays. The absorption via interstellar radiation fields in the plane depend on the Galactic longitude; the dashed lines indicate observations for a source at the GC where the absorption effect is strongest~\cite{Moskalenko:2005ng}. Note that the individual diffuse TeV-PeV $\gamma$-ray limits of the GP are for different emission regions along the GP as indicated in the legend of the plot. The relative size of the ``on-source'' regions of the experimental results are summarized in Fig.~\ref{fig5}. The diffuse flux prediction (only $\pi^0$-decay) for $|b|<5^\circ$ or $|b|<10^\circ$ are lower than the $|b|<2^\circ$ calculation shown in Fig.~\ref{fig4} by about a factor 2 or 3, respectively.

The intensity of the Galactic diffuse emission (including unresolved point source emission and truly diffuse emission) is also expected to vary along the GP. For a uniform source distribution or CR density within the GP (as assumed in our approximation) the flux variation between the Galactic Center to anti-Center is less than 25\% (omitting absorption). For instance, the flux predictions in the inner (outer) Galaxy corresponding to the Tibet limits ({\it cf.}~Figs.~\ref{fig4} and \ref{fig5}) increase (decrease) by 20\% (23\%) compared to the overall average. However, as mentioned earlier, one has to keep in mind that the source distribution should also follow the Galactic arms, bar and bulge. Similar to the observed $\gamma$-ray distribution along the GP this can enhance the neutrino emission in directions with increased local source density.

The Milagro experiment identified a diffuse $\gamma$-ray emission in the GP at 3.5~TeV within $40^\circ<\ell<100^\circ$ and at 15~TeV within $40^\circ <\ell<85^\circ$~\cite{Atkins:2005wu,Abdo:2008if}. The cumulative flux of many sources including SNRs or PWNe may make a significant contribution to the Milagro flux. This is roughly consistent with estimates based on analyses on nearby SNRs and PWNe that have been observed by Cherenkov telescopes like HESS~\cite{Casanova:2007cf}. The neutrino flux from SNRs suggested by Eqs.~(\ref{eq:Jrel}) and (\ref{eq:SNR}) is marginally consistent with diffuse GP $\gamma$-ray measurements. Deeper limits are important to test if young SNRs are responsible for CRs around the knee. Theoretically, if sufficient magnetic amplification in the upstream region is possible, $E_{p,{\rm max}} \simeq E_{\rm kn}$ can be achieved only around $t_{\rm Sed}$. Non-observations of PeV $\gamma$-rays will imply that not all young SNRs in the Sedov phase do accelerate CRs up to the CR knee, which has already been suggested from TeV $\gamma$-ray observations of nearby SNRs~\citep[{\it e.g.},][]{Funk:2010cg}.

The diffuse flux from the propagation of CRs is also consistent with the estimate of Eqs.~(\ref{eq:Jrel}) and (\ref{eq:JCR}) and dominates at lower energies. The extrapolated flux in the GeV-TeV region is consistent with the overall observed flux of GP diffuse GeV-TeV $\gamma$-ray emission by Fermi~\cite{FermiLAT:2012aa} if one considers uncertainties of the local diffuse CR spectrum, leptonic contributions and matter distribution throughout the Galaxy.

More generally, we can see from Fig.~\ref{fig4} that $>100$~TeV $\gamma$-ray limits in the GP are at a comparable level as (or at a slightly lower level than) the diffuse isotropic limits shown in Fig.~\ref{fig2}. Note that the limits and measurements of the diffuse GP flux are obtained after subtracting the isotropic component inferred from an ``off-source'' region with varying size and position in each experimental study. To quantify the systematic uncertainty from the contribution of signal to the background measurement, we can assume that the off-source region is defined as an adjacent region at the same longitudinal range, but with $b_{\rm max}<|b|<2\times b_{\rm max}$. Assuming $b_{\rm max}=2^\circ$ the contribution of diffuse CRs in the off-source region is less than 30\% relative to the on-source region and introduces a relative systematic uncertainty of the upper limits at the same level. An extragalactic diffuse $\gamma$-ray emission in the $>100$~TeV energy range will be strongly suppressed due to photon absorption in the extragalactic background light. 

If all events of the IceCube excess would be associated with the GP at $|b|<2^\circ$ the diffuse GP flux would be about $4\pi/\Delta\Omega_{\rm GP}\simeq29$ times larger than the prediction in Eqs.~(\ref{eq:ICnu}) and (\ref{eq:Jrel}). This is clearly ruled out by the diffuse GP limits shown in Fig.~\ref{fig4}. However, already 4\% of the IceCube excess, {\it i.e.}~about one out of the 28 would correspond to a diffuse GP flux at the same level as the isotropic prediction. The association of the GP emission with the IceCube excess is hence very unlikely. Obviously, statistical fluctuations and the different FoV of $\gamma$-ray observatories are important for a more quantitative estimate, but this doesn't change the general argument.

Deeper PeV $\gamma$-ray observations covering the GP can test the SNR/HNR scenario more solidly, independently of an association with the IceCube excess. In Fig.~\ref{fig4} we show the sensitivity of the air shower arrays HAWC~\cite{DeYoung:2012mj} (3 years), LHAASO~\cite{Zha:2012wp} (1 year) and HiSCORE~\cite{Tluczykont:2012nm} (5 years) and for the proposed Cherenkov Telescope Array (CTA)~\cite{Consortium:2010bc} (100 hours). For CTA we assume a FoV with diameter of $10^\circ$ and $\theta_{\rm PSF}\simeq0.05^\circ$. To account for the limited FoV of these experiments we estimate the upper diffuse limits from the point source (PS) sensitivities $\Phi^{\rm PS}$ (in units of ${\rm GeV}^{-1}{\rm cm}^{-2}~{\rm s}^{-1}$) via $\Phi^{\rm diff} \sim \Phi^{\rm PS}/\sqrt{\Omega_{\rm GP\,\cap\,FoV}\Omega_{\rm PSF}}$ (in units of ${\rm GeV}^{-1}{\rm cm}^{-2}~{\rm s}^{-1}~{\rm sr}^{-1}$) where $\Omega_{\rm GP\,\cap\,FoV}$ is the size of the GP ($|b|<2^\circ$) in the FoV and $\Omega_{\rm PSF} \simeq \pi\theta_{\rm PSF}^2$ is the size of the point-spread function (PSF). For HAWC and LHAASO we have $\Omega_{\rm GP\,\cap\,FoV}\simeq0.3$~sr and assume $\theta_{\rm PSF}\simeq 0.2^\circ$ which gives a correction $\Phi^{\rm PS}/\Phi^{\rm diff} \simeq 3.4\times10^{-3}$~sr in both cases. For HiSCORE we assume a site location at $35^\circ$S with $\Omega_{\rm GP\,\cap\,FoV}\simeq0.2$~sr and also $\theta_{\rm PSF}\simeq 0.2^\circ$. For CTA we assume $\Phi^{\rm PS}/\Phi^{\rm diff} \simeq 1.7\times10^{-4}$~sr. These observatories should be able to provide further constraints on the hadronic emission scenario of SNR/HNR after a few years of observation.

%%%
\begin{figure*}[t]
\includegraphics[height=0.38\linewidth]{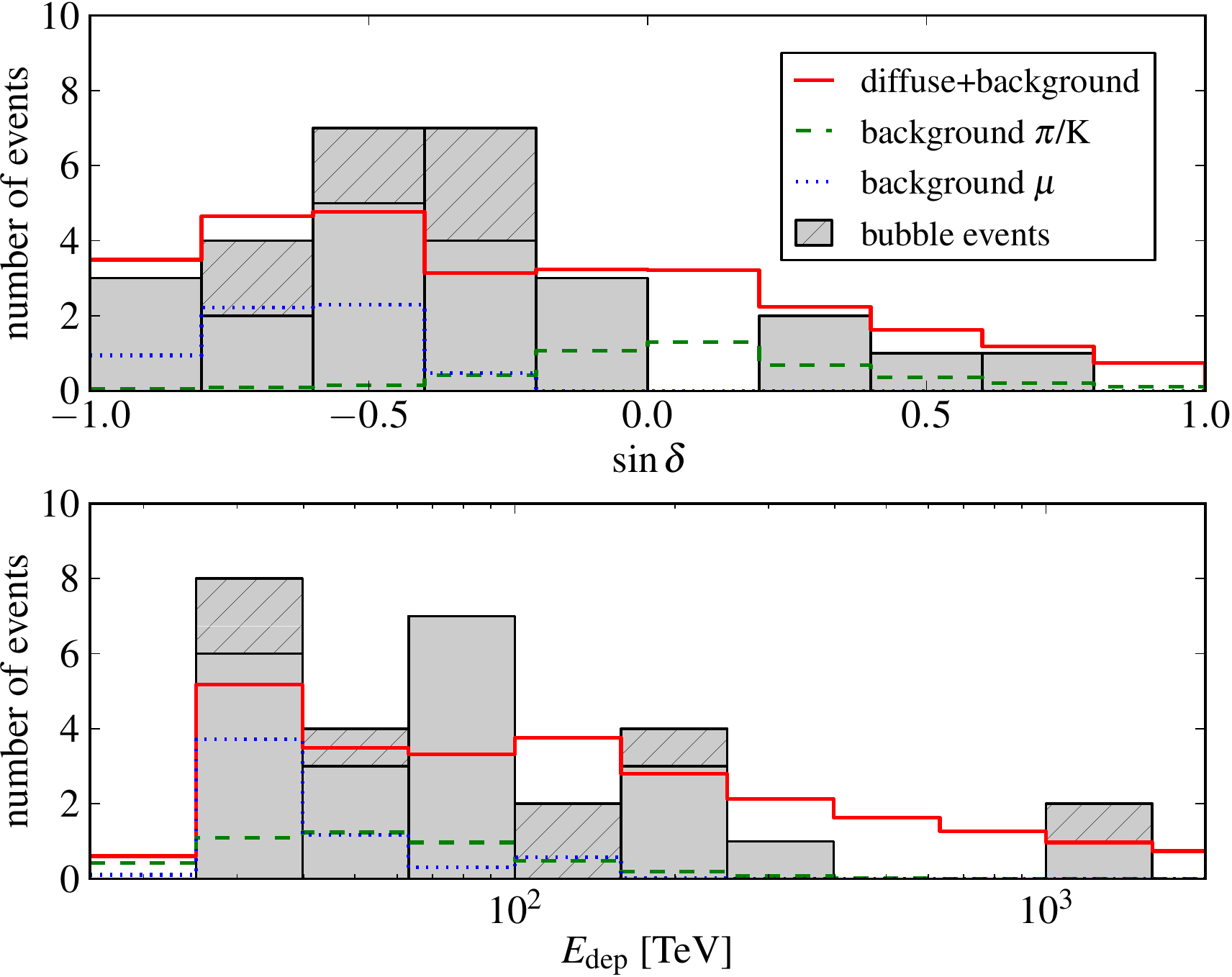}\hfill
\includegraphics[height=0.38\linewidth]{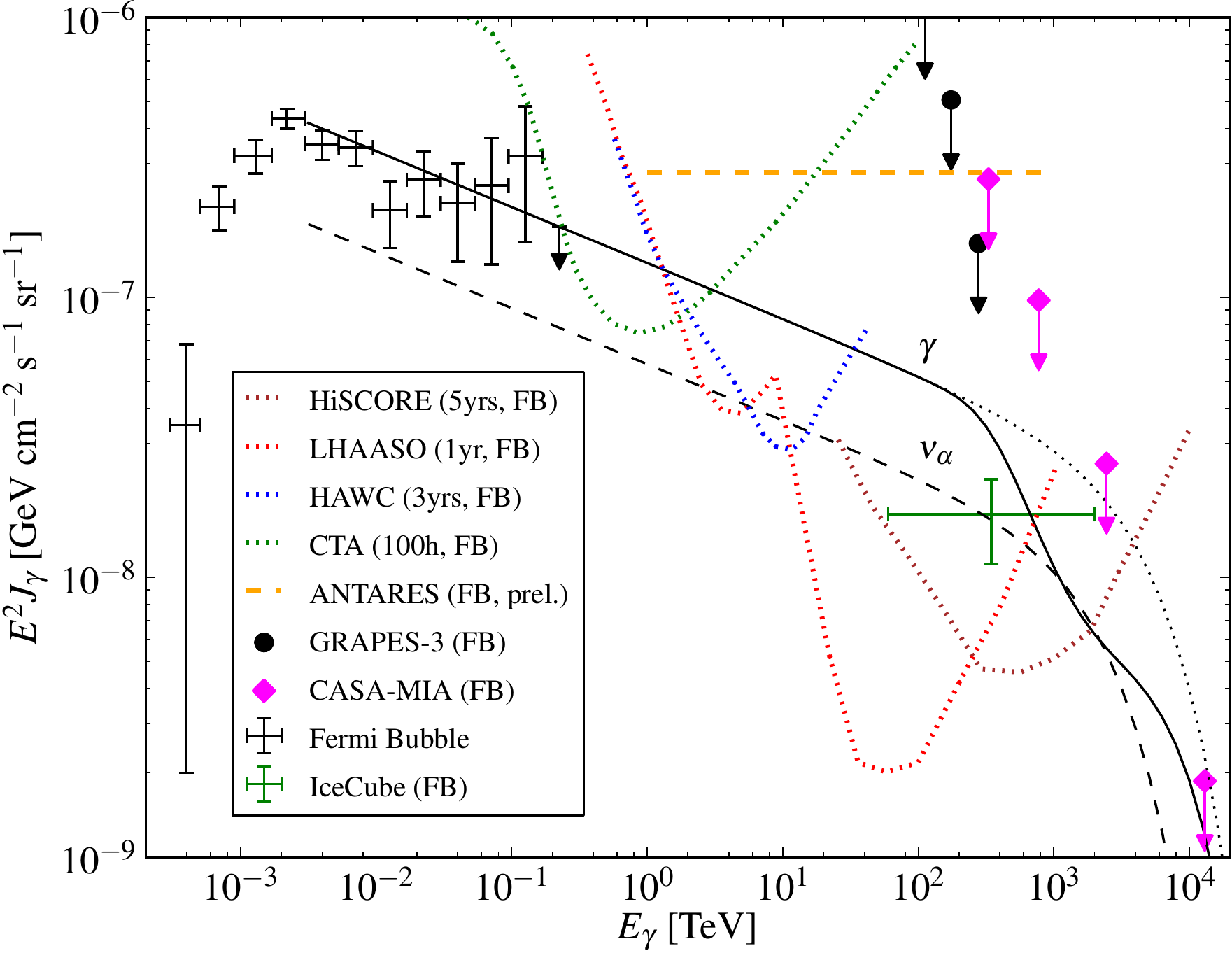}
\caption[]{{\bf Left:} Histogram of the event distribution in declination (top) and deposited energy (bottom). The hatched area shows the contribution of the seven events in the extended GC region with a possible association with the FBs. The lines shows the expected background from atmospheric muons (dotted), conventional atmospheric neutrinos (dashed) and the sum of these backgrounds and the best-fit diffuse flux (solid) from Ref.~\cite{Aartsen:2013pza}. {\bf Right:} The diffuse flux from the FB in comparison with diffuse $\gamma$-ray limits in the 0.1-1~PeV range corrected for the overlap of the FoV with the FB region. The horizontal dashed line is a preliminary upper limit from ANTARES on the per flavor neutrino flux of the FB~\cite{Mangano:2013dta}. The green point indicates the equivalent diffuse flux from the FB of $1.4J^{\rm IC}_{\nu_\alpha}$ (see main text). The dotted (solid) line shows a possible intrinsic (absorbed) $\gamma$-ray emission from the FB with a spectral index $\Gamma=2.2$ and exponential cutoff at 6~PeV according to Eq.~(\ref{eq:FBnu}). The corresponding neutrino flux (per flavor) is shown as a dashed line. We also show estimates of the sensitivity of CTA (green dotted), HAWC (blue dotted), LHAASO (red dotted) and HiSCORE (brown dotted) w.r.t.~the diffuse TeV-PeV $\gamma$-ray emission in the FBs.}\label{fig6}
\end{figure*}
%%%

\subsubsection{Fermi Bubbles}

{\it Fermi Bubbles} (FBs)~\cite{Su:2010qj} are hard and uniform emission regions of 1-100~GeV $\gamma$-rays detected by Fermi extending above and below the GP to a distance of $\pm10$kpc. It has been suggested that this emission is due to hadronuclear interactions of CRs that are possibly accelerated by star-burst driven winds and convected from the GC region over time scales of the order of several Gyrs~\cite{Crocker:2010dg}. As discussed in the previous section, the $pp$ reaction will also provide a hard spectrum of neutrinos in the FBs~\cite{Crocker:2010dg,Lunardini:2011br,Cholis:2012fr}. Note, however, there is also the leptonic emission model for the FBs, in which associated neutrinos are not expected~\cite{Mertsch:2011es}.

If CRs are injected with a luminosity $L_p\sim{10}^{39}~{\rm erg}~{\rm s}^{-1}$~\cite{Lacki:2013zsa} over several billions of years it is expected that the proton population in the FBs reaches a quasi-steady state~\cite{Crocker:2010dg}. Similar to the discussion in the previous section we can estimate the proton spectral injection rate as$E_p^2Q_p(E_p) \simeq L_p(E_p/E_{p,\rm min})^{2-\Gamma}/\mathcal{R}_0$ and the neutrino spectral emissivity is given in steady state as $E_\nu^2Q_{\nu_\alpha}(E_\nu)\simeq(1/6)E_p^2Q_p(E_p)$ with $E_\nu \simeq 0.05E_p$. In reality, protons will lose their energies simultaneously via ionization and adiabatic losses~\cite{Crocker:2010dg} and hence the expected neutrino emissivity will be somewhat smaller than this estimate. Then, with $\Delta\Omega_{\rm FB}\simeq1.2$~sr, we can estimate the as
\begin{multline}\label{eq:FBnu}
E_\nu^2 J^{\rm FB}_{\nu_\alpha} \sim \frac{1}{\Delta\Omega_{\rm FB}}\frac{E_\nu^2Q_{\nu_\alpha}}{4\pi r^2_{\rm FB}}\\
\simeq  7.2\times10^{-6}~{\rm GeV}~{\rm cm}^{-2}~{\rm s}^{-1}~{\rm sr}^{-1}\frac{1}{\mathcal{R}_0}\left(\frac{E_{\nu}}{E_{\nu,{\rm min}}}\right)^{2-\Gamma}\\
\times L_{p, 39} r_{\rm FB, 1}^{-2}\,,
\end{multline}
where $r_{\rm FB}$ is the fiducial distance to the FB. In the following we will use $r_{\rm FB}=8.5$~kpc, $\Gamma=2.2$ ($\mathcal{R}_0 \simeq 4.8$) and $L_p\simeq 2\times10^{38}$~erg/s which is consistent with the GeV $\gamma$-ray flux of $\sim4\times{10}^{-7}~{\rm GeV}~{\rm cm}^{-2}~{\rm s}^{-1}~{\rm sr}^{-1}$. 

The accumulation of seven cascade events within about $30^\circ$ off the GC includes the hot-spot in IceCube's event cluster search with a trial-corrected significance of 8\%~(see Fig.~\ref{fig1}). The histogram in the left panel of Fig.~\ref{fig6} shows the distribution of these seven events in declination and detected energy. The declination distribution of the reduced sample of 21 events follows the isotropic distribution more closely, as can be seen in the top panel. On the other hand, there are no noticeable qualitative changes of the energy distribution of the reduced sample shown in the lower panel. This suggests that a combined fit by the FBs+isotropic neutrino flux might provide a better description of the data. Note, that the deposited energy is only a lower bound on the neutrino energy. In the case of cascades from neutral current interactions an average fraction of $70-80$\% is carried away by the invisible neutrino and the energy deposited by the muons depend on the track length and can be smaller by orders of magnitude. The apparent gap of events the energy distribution shown in the lower histogram of the left panel in Fig.~\ref{fig6} might be due to this effect.

We estimate the per-flavor flux of the FBs via the contribution of all weighted events to two spherical regions above and below the GP with a radius of $25^\circ$ which gives $n_{\rm FB}\simeq 3.6$. Using Eqs.~(\ref{eq:ICnu}) and (\ref{eq:mod}) we arrive at $J^{\rm FB}_{\nu_\alpha}(E_\nu) \simeq  2.2 (1.4) J^{\rm IC}_{\nu_\alpha}(E_\nu)$ for $E_\nu$ in the IceCube energy range and including (excluding) the isotropic background of the rest of the IceCube excess. Since the spectral index of this flux as well as the neutrino energy range is not well determined we show the corresponding neutrino flux of the FBs (without background) as one data point in the right panel of Fig.~\ref{fig6}. We also show an estimate of the diffuse limits from CASA-MIA and GRAPES-3 which have a small overlap with the Northern FB. We correct the limits by the factor $\sqrt{\Omega_{\rm FoV}/\Omega_{\rm FB\,\cap\,FoV}}$, where $\Omega_{\rm FoV}$ is the size of the observatory's field of view (FoV) and $\Omega_{\rm FB\,\cap\,FoV}$ the size of its intersection with the FBs. For CASA-MIA and GRAPES-3 the intersection has a size of $0.44$~sr and $0.30$~sr, respectively, resulting in a correction of the upper diffuse limit by factors $4.4$ and $4.2$.

We also indicate that possible neutrino and $\gamma$-ray emissions from the hadronic scenario of the FBs are consistent with neutrino and $\gamma$-ray observations. We assume a reference $\gamma$-ray spectrum with spectral index $\Gamma\simeq 2.2$ and exponential cutoff at 6~PeV. This would require a CR population in the FBs with an exponential cutoff at 60~PeV, well above the CR \textit{knee}. In fact, the FBs have also been suggested as possible accelerators of CRs above the CR \textit{knee}~\cite{Lacki:2013zsa,Cheng:2011tx}. The horizontal dashed line in the plot indicates a preliminary diffuse neutrino limit of the ANTARES Collaboration~\cite{Mangano:2013dta}. Located in the Northern Hemisphere, ANTARES can search for neutrinos of most of the FBs with the traditional muon neutrino detection channel of up-going tracks. The present limit is consistent with IceCube's observation of seven events from the FB region. The proposed future Mediterranean telescope KM3NET is expected to improve this limit by an order of magnitude after one year of observation~\cite{Adrian-Martinez:2012qpa}. 

In addition, combining deeper PeV $\gamma$-ray observations covering the IceCube sky should enable us to test this scenario solidly. We indicate in the right panel of Fig.~\ref{fig6} the sensitivity of CTA, HAWC, LHAASO and HiSCORE to the diffuse emission of the FBs. Again, for CTA we assume a FoV with diameter of $10^\circ$ and PSF with $\theta_{\rm PSF}\simeq0.05^\circ$. If the FoV is contained in the FB (depending on the final location of the observatory) this gives a correction $\Phi^{\rm PS}/\Phi^{\rm diff}\simeq2.4\times10^{-3}$~sr. This estimate may be optimistic since the search for extended emission with Imaging Atmospheric Cherenkov Telescopes like CTA requires an ``edge''-like $\gamma$-ray emission. Such an edge with about $2^\circ$ width is in fact suggested by the Fermi data~\cite{Su:2010qj}, but more sophisticated studies are needed. We estimate the size of the overlap region $\Omega_{\rm FB\,\cap\,FoV}$ of HAWC, LHAASO and HiSCORE as $0.7$~sr, $0.5$~sr and $1.0$~sr, respectively. This gives a relative correction $\Phi^{\rm PS}/\Phi^{\rm diff}$ for $\theta_{\rm PSF}\simeq0.2^\circ$ of $5.3\times10^{-3}$~sr, $4.2\times10^{-3}$~sr and $6.2\times10^{-3}$~sr, respectively. Again, this can only be considered an estimate since the experimental acceptance drops towards the edge of the FoV. Nevertheless, all observatories have the possibility to test the hadronic emission model of the FBs with $\Gamma\simeq2.2$ after a few years of observation.

\section{Conclusions}

The IceCube excess of 28 neutrino events in the TeV-PeV energy region opens an exciting new window into the non-thermal Universe. It is an open question if the observed flux is a nearly isotropic emission that would naturally originate from an extragalactic source distribution, or if the data hint to substructures that could point to an extended region around the GC or the GP. 

In this paper we have studied in detail how the TeV-PeV $\gamma$-rays produced via the same hadronic CR interactions responsible for the neutrino emission can identify or exclude Galactic contributions. We have summarized upper limits on isotropic and non-isotropic diffuse $\gamma$-ray emission. We point out that TeV and PeV $\gamma$-ray upper limits placed by Fermi and EAS detectors already give us intriguing constraints on the possibility that the IceCube excess has an (quasi-)isotropic Galactic origin. Such nearly isotropic emission could be produced by CR interactions with circumgalactic material in the Galactic halo or PeV DM decay. However, more than half of the IceCube events originate in a region of the sky which is not constrained by present limits due to the limited field of view of extended air shower observatories, which are all located in the Northern Hemisphere except for IceCube/IceTop.

We also discussed scenarios of extended neutrino/$\gamma$-ray emission of clusters of Galactic sources or extended Galactic sources on top of an extragalactic diffuse flux. To identify weak clustering in IceCube's event distribution we weighted the IceCube events based on an average signal-to-background distribution in declination and observed energy. We identified maximal fluctuations from an isotropic distribution in the GC region and close to the GP at $\ell\simeq240^\circ$. Presently, these fluctuations are {\it not} statistically significant, but they serve as a motivation for theoretical speculations about possible emission regions in combination with present $\gamma$-ray observations. With this in mind, we studied the expected neutrino and $\gamma$-ray fluxes from the non-isotropic diffuse emission in the GP and from the FB feature observed in GeV $\gamma$-rays. 

For the GP emission we studied the soft diffuse emission from CR interactions with interstellar gas and the hard emission from unresolved CR sources which we model via SNRs/PWNe as well as HNRs. We show that both scenarios are marginally consistent with upper limits on the diffuse emission in the GP. However, the former case is an unlikely contribution to the IceCube excess due to the expected soft and uniform emission close to the GP.  The latter case might contribute to the fluctuation in the GP at $\ell\simeq240^\circ$ which is close to the direction of the Local Arm possibly containing close-by CR sources. Given that SNRs accelerate CRs up to the knee, HNRs can accelerate CRs up to $20-30$~PeV, leading to PeV neutrinos. Future PeV $\gamma$-ray observatories can further test the contribution from both SNRs and HNRs to the IceCube excess.

For the FBs we can estimate the hadronic $\gamma$-ray and neutrino emission via an extrapolation of the observed $1-100$~GeV emission. We could show that the contribution of the IceCube excess to the FB region is equivalent to a $\gamma$-ray flux that follows the extrapolation with a power index $\Gamma\simeq2.2$ assuming that Galactic sources accelerate CRs to the required energies. The future $\gamma$-ray observatories HAWC, LHAASO and HiSCORE and possibly CTA can test this hard hadronic emission independently.

We also indicated that exotic origins of the IceCube excess in the from of DM decay in the Galactic halo can be tested by their PeV $\gamma$-ray emission. We discussed a specific model of a scalar DM particle decaying into two Higgs which is marginally consistent with the non-observation of PeV $\gamma$-rays. However, different DM scenarios with line features or extended decay channels, {\it e.g.}~$X \to\tau^+\tau^-$ can lead to an increased PeV $\gamma$-ray emission that can already be excluded by diffuse TeV-PeV $\gamma$-ray limits.

{\it While we were waiting for the permission of the IceCube Collaboration to proceed with this preprint we became aware of Ref.~\cite{Razzaque:2013uoa} also pointing out a possible origin of the GC events in the FBs. Preliminary results of this analysis has been presented prior to that preprint~\cite{AhlersVLVnT,AhlersTeVPA}.}
\vspace{0.4cm}
\acknowledgments
We would like to thank John Beacom, Francis Halzen, Aya Ishihara, Albrecht Karle, Claudio Kopper, Naoko Kurahashi, Christian Spiering, Floyd Stecker, Stefan Westerhoff and Nathan Whitehorn for very helpful discussions and comments. We also thank the anonymous referee for pointing out the tension between Milagro $\gamma$-ray observations and GRAPES-3 upper limits in the inner GP. MA acknowledges support by the {\it Wisconsin IceCube Particle Astrophysics Center} (WIPAC) and the U.S. National Science Foundation (NSF) under grants OPP-0236449 and PHY-0236449. KM is supported by NASA through Hubble Fellowship grant No.~51310.01 awarded by the Space Telescope Science Institute, which is operated by the Association of Universities for Research in Astronomy, Inc., for NASA, under contract NAS 5-26555.

\end{document}